\documentclass[twocolumn,amsmath,amssymb,aps,prd,nofootinbib]{revtex4-2}

\usepackage{graphicx}
\usepackage{multirow}

\begin{document}

\title{Coupled quintessence with a generalized interaction term}

\author{Robertus Potting$^{1,2}$}

\email{rpotting@ualg.pt}

\author{and Paulo M.~S\'a$^{1,3}$}

\email{pmsa@ualg.pt}

\affiliation{
$^1$Departamento de F\'\i sica, Faculdade de Ci\^encias e Tecnologia, 
Universidade do Algarve, 8005-139 Faro, Portugal}

\affiliation{
$^2$Centro de Astrof\'\i sica e Gravita\c c\~ao,
Instituto Superior T\'ecnico, Universidade de Lisboa, 
Avenida Rovisco Pais, 1049-001 Lisbon, Portugal}

\affiliation{
$^3$Instituto de Astrof\'\i sica e Ci\^encias do Espa\c co,
Faculdade de Ci\^encias da Universidade de Lisboa, Edif\'\i cio C8,
Campo Grande, 1749-016 Lisbon, Portugal}

\begin{abstract}
We investigate a cosmological model in which dark energy,
represented by a quintessential scalar field, 
is directly coupled to a dark-matter perfect fluid.
We are interested in solutions of cosmological relevance,
namely those for which a dark-matter-dominated era long enough
to allow for structure formation is followed by an era of accelerated
expansion driven by dark energy.
For the coupling between these two dark components of the Universe,
we choose forms that generalize the one most commonly used in the literature.
Resorting to powerful methods of qualitative analysis of dynamical systems,
we show that, for certain generalized forms of the coupling,
final states of our coupled quintessential model correspond to 
solutions in which the evolution of the Universe is completely dominated by
dark energy. 
In this case, there are no scaling solutions.
Interestingly, however, for certain values of a relevant parameter, during
the approach to the final state of evolution, the cosmological
parameters change so slowly that, for all practical purposes, the
solution appears to be stuck in a state corresponding to a scaling solution.
\end{abstract}

\date{17 May, 2022}

\maketitle

\section{Introduction}

Ever since the discovery of the accelerated expansion of the Universe
\cite{riess-1998,perlmutter-1999} it has been clear that there is a need 
for a consistent theoretical model to explain the origin of this acceleration. 
The simplest candidate model involves a cosmological constant, whose energy
density remains unchanged throughout the evolution of the Universe.
Even though consistent with current observational data, the cosmological
constant is problematic, as theoretical predictions of its value vastly 
exceed the observed one \cite{weinberg-1989,martin-2012}.
An appealing alternative approach that has been proposed is that the role
of dark energy be played by quintessence: a dynamical scalar field,
whose potential energy starts to dominate the evolution of the
Universe at a relatively late stage of its evolution,
giving rise to a period of cosmic acceleration,
in a way similar to primordial inflation \cite{caldwell-1998}
(for reviews on dynamical dark energy see, e.g., 
\cite{copeland-2006,bamba-2012}).

Another observed phenomenon whose nature still remains unknown is
dark matter \cite{bertone-2018}.
In the above context it is appealing to identify it to a scalar field as well.
This opens the possibility to unify two seemingly disparate phenomena within
the same theoretical framework. Such an approach was carried out, 
for instance, in Refs.~\cite{sa-2020b,sa-2021}, where dark energy and dark
matter were unified within a two-scalar-field cosmological model.

In the current work we will treat dark matter as a perfect fluid.
However, note that this fluid can in principle be modelled by a scalar field.
We will represent dark energy by a quintessencial scalar field
with an exponential potential, which is a common choice in much of the
literature that can be motivated from string theory.
The second key assumption we make is that the dark energy and dark matter
sectors are coupled, that is, there is an energy transfer between the two.
Note that, as long as we are only able to probe these dark components
through their gravitational effects, one cannot exclude
the possibility that there are non-gravitational interactions between them,
and thus one is naturally led to consider cosmological theories
involving interactions between the two.

There have been quite a few studies with different forms of the
quintessence to dark matter coupling considered in the literature
(for a review, see,  for instance, Refs.~\cite{bolotin-2015,wang-2016}).
A common choice has been $Q \propto \rho_\texttt{DM} \dot\phi$,
where $\phi$ is the quintessence field and $\rho_\texttt{DM}$ the dark 
matter energy density
\cite{amendola-1999,holden-2000,billyard-2000,amendola-2000,
tocchini-valentini-2002,gumjudpai-2005,boehmer-2008,tzanni-2014,singh-2016}.
In this case, the evolution equations admit cosmological
solutions with late-time accelerated expansion of two types:
(i) the dark-energy scalar field dominates the dynamics of the universe
and the energy density of the dark-matter fluid vanishes rapidly;
(ii) the so-called scaling solutions for which the ratio between the energy
densities of dark matter and dark energy converge asymptotically to a 
nonzero value.
The latter is of physical interest as a possible way to address the
cosmological coincidence problem.
As it turns out, however, for those scaling solutions the accelerated expansion
is not preceded by a  matter-dominated era long enough to allow  for the
observed structure formation \cite{bahamonde-2018}
which rather limits the physical relevance of these solutions.

In order to remedy this problem,
we will consider in this work a class of couplings of the form
$Q \propto \rho_\texttt{DM} C(\phi) \dot\phi$,
where $C(\phi)$ is a function of the scalar field.
As we will show in section \ref{Sec-interaction-term},
this particular form of the interaction term can be motivated
by a variational approach to coupled quintessence \cite{boehmer-2015,boehmer-2015b}.
Our goal is to obtain solutions with a dark-matter-dominated era long enough
to allow for structure formation, followed by an era of accelerated
expansion driven by dark energy.
First we take $C(\phi)$ to be a small harmonic fluctuation around the constant
value one and investigate the fate of the scaling solutions
mentioned above.
Next, we consider the case in which $C(\phi)$ is an unbounded function;
explicitly, we take $C(\phi)\propto \phi^n$, with $n=1,2,3,\ldots$
In this work we use methods of qualitative analysis
of dynamical systems,
which have been applied with great
success to astrophysical and cosmological problems for several decades.
It turns out that, at least qualitatively,
all cases with $n$ equal to a positive integer behave in a similar way.
For this reason, only the case $n=1$ is worked out in full detail.

This article is organized as follows.
In section \ref{Sec-interaction-term} we motivate the coupling between
the quintessential scalar field and the dark-matter fluid.
In section \ref{Sec-standardCQ}, scaling solutions in
the standard coupled quintessential cosmological model
are reviewed.
In section \ref{Sec-perturbed} a model in which $C(\phi)$ includes a small 
harmonic fluctuation is shown to exibit perturbed scaling solutions.
Next, in section \ref{Sec-CQgeneralized}, $C(\phi)$  is taken to be
proportional
to $\phi^n$.
The resulting solutions are analyzed in section \ref{Sec-DSsolutions},
with a special focus on the solutions of cosmological relevance
in section \ref{Sec-cosmological}.
Finally, we present our conclusions in section \ref{Sec-conclusions}.

\section{The interaction term between dark energy and dark matter\label{Sec-interaction-term}}

It is a standard procedure in coupled quintessential models
to introduce the interaction between dark matter and dark energy at the
level of the cosmological field equations.
More specifically,
for a flat Friedman--Lema\^{\i}tre--Robertson--Walker (FLRW)
universe\footnote{Since the current cosmological measurements constrain
the present-time value of the curvature density parameter $\Omega_k$
to be very small \cite{ade-2016},
we make the simplifying assumption of a spatially flat Universe.},
the evolution equations for coupled quintessence are given
by\footnote{In this article we adopt units with
$c=\hbar=1$ and use the notation $\kappa\equiv \sqrt{8\pi G}$,
where $G$ is the gravitational constant.} \cite{wetterich-1995}
\begin{gather}
  \label{ddotphi 1}
\ddot{\phi} + 3H\dot{\phi} +\frac{\partial V}{\partial\phi}
= \frac{Q}{\dot{\phi}},\\
  \label{dotrho}
\dot{\rho}_\texttt{DM}+3H\rho_\texttt{DM} = -Q,\\
  \label{dotH}
\dot{H} = -\frac{\kappa^2}{2}  \left(\dot{\phi}^2 + \rho_\texttt{DM} \right),\\
H^2 = \frac{\kappa^2}{3} \left( \frac{\dot{\phi}^2}{2} + V 
+\rho_\texttt{DM} \right),
  \label{Friedmann 1}
\end{gather}
where $\phi$ is the quintessential dark-energy scalar field subject to the
potential
\begin{equation}
V(\phi) = V_a e^{-\mu\kappa\phi},
 \label{Potential}
\end{equation}
with constants $V_a > 0$ and $\mu$ of mass dimension 4 and 0, respectively,
$\rho_\texttt{DM}$ is the energy density of a pressureless dark-matter fluid,
$Q$ the interaction term between dark energy and dark matter,
and $H = \dot{a}/a$ the Hubble parameter.
Overdots denote a derivative with respect to time~$t$.

The introduction of the interaction term $Q$ in the above equations
derives from the addition of a ``coupling current $Q_\mu$" in the
conservation equations of the dark-matter and dark-energy components,
namely,
\begin{equation}
\nabla^\mu T_{\mu\nu}^\texttt{(DE)} = Q_\nu
\quad {\rm and} \quad
\nabla^\mu T_{\mu\nu}^\texttt{(DM)} = -Q_\nu,
\end{equation}
where $T_{\mu\nu}^\texttt{(DE)}$ and $T_{\mu\nu}^\texttt{(DM)}$ are the
energy-momentum tensors of dark energy and dark matter, respectively,
satisfying the Einstein field equations
\begin{equation}
G_{\mu\nu}=\kappa^2 \left(
T_{\mu\nu}^\texttt{(DE)} + T_{\mu\nu}^\texttt{(DM)} \right).
\end{equation}
In the FLRW cosmological scenario considered in this paper only
the time component of $Q_\mu$ is nonvanishing,
which can be identified with the coupling $Q$.
The dependence of $Q$ on $\rho_\texttt{DM}$, $\phi$, and their derivatives is
\textit{a priori} undetermined and amounts to a phenomenological assumption.
A variety of choices for the form of $Q$ have been made by different authors,
the most common being $Q\propto\rho_\texttt{DM} \dot{\phi}$.

In a variational approach, a coupling between a dark-energy scalar field
and dark matter can be introduced already at the Lagrangian level.
Such an approach has been applied to quintessence
by B\"ohmer and collaborators \cite{boehmer-2015,boehmer-2015b}.
Considering algebraic couplings between the scalar field and the matter fluid,
as well as derivative couplings, they have shown that the
Lagrangian formulation can be mapped back into the standard relativistic
approach mentioned above (or, equivalently, every standard coupled
quintessential model can be derived from the proposed Lagrangian formulation).

The relationship between the variational and standard approaches is given by 
\begin{gather}
\label{Q_variational}
Q = -\frac{\partial \rho_\texttt{int}}{\partial\phi} \dot{\phi},
\\
\rho_\texttt{DM} = \rho + \rho_\texttt{int},
\\
p_\texttt{DM} = p + p_\texttt{int} = 0,
\end{gather}
where $\rho$ ($p$) and $\rho_\texttt{int}$
($p_\texttt{int}$) are the energy densities (pressures) of the matter fluid
and the interaction sector, respectively, in the variational approach, while
$\rho_\texttt{DM}$ ($p_\texttt{DM}$) is the energy density (pressure)
of the dark-matter fluid in the standard approach.

An important conclusion we can draw from Eq.~(\ref{Q_variational}) is that
the coupling $Q$ is proportional to $\dot{\phi}$.
Indeed, it follows that the interaction term has the form
\begin{equation}
Q = A(\rho,\phi)\dot{\phi},
\label{Q_general}
\end{equation}
where $A(\rho,\phi)$ is an \textit{a priori} arbitrary function
of $\rho$ and $\phi$, such that
\begin{equation}
 \frac{\partial \rho_\texttt{int}}{\partial\phi} = -A(\rho,\phi)\>.
\end{equation}

It is easy to see that the choice
\begin{equation}
\rho_\texttt{int} = B(\rho) e^{-\kappa\beta\int C(\phi)d\phi} - \rho,
\end{equation}
where $\beta$ is a dimensionless constant and $B(\rho)$ and $C(\phi)$ are
arbitrary functions of $\rho$ and $\phi$, respectively, leads to an 
interaction term in the variational approach of the form
\begin{align}
Q & = \kappa\beta B(\rho) e^{-\kappa\beta\int C(\phi)d\phi} C(\phi) \dot{\phi}
 \nonumber \\
  & = \kappa\beta (\rho+\rho_\texttt{int}) C(\phi) \dot{\phi},
\end{align}
which corresponds, in the standard approach, to
\begin{equation}
Q = \kappa\beta\rho_\texttt{DM} C(\phi) \dot{\phi}.
  \label{Q_Cphi}
\end{equation}
This is the form of the interaction term that will be adopted
in this paper. As we will show below,
for $C(\phi)\propto\phi^n$ ($n=1,2,3,\dots$), 
strictly there are no scaling solutions in the model
defined by Eqs.~(\ref{ddotphi 1})--(\ref{Potential}) and (\ref{Q_Cphi}).
However, for certain values of a relevant parameter,
there are solutions that, during the approach to the final state of
cosmic evolution, behave, for all practical purposes, as accelerated
scaling solutions. Furthermore, these solutions are preceded by an
era of dark-matter domination.

To conclude this section, we point out that coupled dark-energy
cosmological models with an interaction term given by Eq.~(\ref{Q_Cphi})
have been studied in different contexts.

Within $k$-essence, the most general Lagrangian that exhibits scaling
solutions  was found, but, similarly to the standard quintessence,
it was not possible to build a cosmological model with a correct sequence
of cosmic epochs, namely, a long enough era of matter domination followed
by a present era of accelerated expansion \cite{amendola-2006}.

In cubic-order Horndeski theories and in degenerate higher-order
scalar-tensor theories, for which gravitational waves 
propagate at the speed of light, the existence of scaling solutions
constrains the coupling  to be of the form 
$C(\phi) \propto 1/(c_1\phi+c_2)$, where $c_1$ and $c_2$ are 
constants \cite{frusciante-2018,frusciante-2019}.
In the case of a constant coupling ($c_1=0$), it is possible to find
cosmological solutions for which an era of accelerated expansion
is preceded by an era of matter domination
\cite{frusciante-2018,frusciante-2019}.

\section{Scaling solutions in standard coupled quintessence\label{Sec-standardCQ}}

A common choice in the literature 
(see Ref.~\cite{bahamonde-2018} and references therein)
is to take $C(\phi)=1$, yielding
\begin{equation}
Q = \kappa\beta\rho_\texttt{DM} \dot{\phi}.
\label{Q0}
\end{equation}
An interaction term of this type has been shown to arise in Brans–Dicke
theory and also in more general non-minimally-coupled gravitational
theories (see, e.g., Ref.~\cite{amendola-2000}).

For the interaction term (\ref{Q0}),
the dynamical system describing the evolution of the coupled quintessential
cosmological model becomes particularly simple.
Indeed, using the dimensionless variables \cite{copeland-1998}
\begin{equation}
x = \frac{\kappa}{\sqrt6 H} \dot{\phi}
  \quad \mbox{and} \quad
y = \frac{\kappa}{\sqrt3 H} \sqrt{V_a e^{-\mu\kappa\phi}}
 \label{xy}
\end{equation}
and a new time variable $\tau$, defined by
\begin{equation}
  \frac{d\tau}{dt}=H \quad \Rightarrow \quad \tau = \ln a,
  \label{tau-t}
\end{equation}
the system of equations~(\ref{ddotphi 1})--(\ref{Friedmann 1}) 
yields the two-dimensional dynamical system
\begin{subequations}
 \label{DS0}
\begin{align}
 x_\tau = & -3x + \frac{\sqrt{6}}{2} \mu y^2 + \frac32 x(1+x^2-y^2)
 \nonumber \\
 & +\frac{\sqrt{6}}{2} \beta (1-x^2-y^2),  \\
 y_\tau = &-\frac{\sqrt{6}}{2} \mu x y + \frac32 y(1+x^2-y^2),
\end{align}
\end{subequations}
where the subscript $\tau$ denotes a derivative with respect to this variable.
The simplicity of this dynamical system stems from the fact that the
interaction term given by Eq.~(\ref{Q0}) can be expressed only in terms
of the variables $x$ and $y$, making it unnecessary to introduce an extra 
variable and the corresponding extra equation.

From the Friedmann equation (\ref{Friedmann 1}) follows
\begin{equation}
 x^2 + y^2 + \Omega_\texttt{DM} = 1, \qquad
 \Omega_\texttt{DM} \equiv \rho_\texttt{DM} \frac{\kappa^2}{3H^2},
   \label{friedmann xy}
\end{equation}
implying that $x^2 + y^2 \leq 1$, since the density parameter
$\Omega_\texttt{DM}$
is, by definition, non-negative. Furthermore, since the dynamical
system~(\ref{DS0}) does not permit orbits that cross the boundary $y=0$,
we restrict the phase space to the region $y\geq0$, which corresponds to
expanding universes. Altogether, the phase space of the dynamical
system~(\ref{DS0}) is the upper half of the unit disk centered at the 
origin $\{(x,y)| x^2 + y^2 \leq 1, y \geq 0\}$.

Without any loss of generality, we can assume the parameter $\mu$ to be
non-negative and the parameter $\beta$ to take any value.
Indeed, since the dynamical system~(\ref{DS0}) is invariant under the 
transformation $x\rightarrow-x$, $\mu\rightarrow-\mu$, 
and $\beta\rightarrow-\beta$, solutions for negative values of $\mu$
can be obtained straightforwardly from solutions for positive values
of $\mu$, provided a reflection over $x$ is performed, as well as a change
of sign of the parameter $\beta$ (of course, we could also have
chosen $\beta$ to be non-negative and $\mu$ to take any value).
Therefore, the parameter space of this model is the half-plane
$\{(\mu,\beta)| \mu\geq0\}$.

The dynamical system~(\ref{DS0}) has five critical points 
\cite{amendola-1999}, two of which correspond to scaling
solutions, namely,
\begin{equation}
 (x_A,y_A) = \left(\frac{2\beta}{\sqrt6},0 \right), 
	\label{CP-A}
\end{equation}	
\begin{equation}
 (x_B,y_B) = \left( \sqrt{\frac32}\frac{1}{\mu - \beta},
 \frac{\sqrt{2\beta^2 - 2\beta\mu + 3}}{\sqrt2|\mu - \beta|} \right).
	\label{CP-B}
\end{equation}

The latter scaling solution, which for certain values of the parameters
$\beta$ and $\mu$ corresponds to a state of accelerated 
expansion\footnote{For the first critical point, the effective 
equation-of-state parameter is $w_{\rm eff}=x_A^2-y_A^2=2\beta^2/3$,
which is always non-negative, while for the second critical point
$w_{\rm eff}=x_B^2-y_B^2=\beta/(\mu-\beta)$, which is smaller than $-1/3$ for
$2\beta+\mu<0$.}, has attracted a lot of attention
\cite{amendola-1999,holden-2000,billyard-2000,amendola-2000,
tocchini-valentini-2002},
because it offers the possibility of solving the cosmological coincidence
problem, i.e., to explain, without fine tuning of the initial conditions,
why the energy densities of dark matter and dark energy
are of the same order of magnitude at the present time.
Unfortunately, such accelerated scaling solutions are not preceded
by a long enough matter-dominated era \cite{bahamonde-2018},
a circumstance that led to a decline in interest in these solutions.
Recently, a two-scalar-field cosmological model was constructed
with accelerated scaling solutions which are preceded by a long enough
era of matter domination; however, the stage of accelerated
expansion, encompassing the present epoch, is temporary \cite{sa-2021}. 

\begin{figure}[t]
 \includegraphics{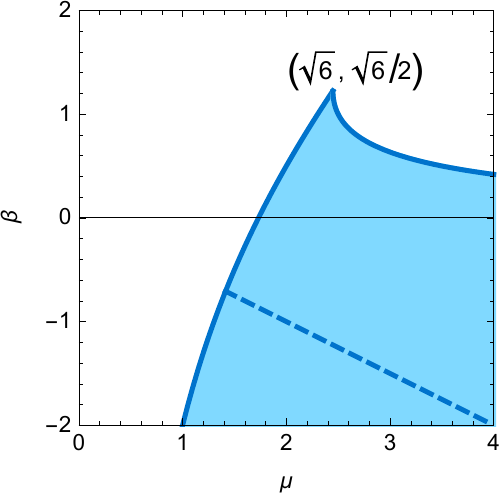}
\caption{\label{Fig-ParameterSpace-CP} The critical point 
$(x_B,y_B)$, given by Eq.~(\ref{CP-B}),
exists in the blue (shaded)
region of the parameter space $(\mu,\beta)$.
Below the dashed line the expansion is accelerated.}
\end{figure}

Let us point out that the critical point given by Eq.~(\ref{CP-B})
only exists for values of $\mu$ and $\beta$ satisfying
the conditions
\begin{equation}
	 0<\mu\leq \sqrt6 \quad \wedge \quad
	\beta \leq \frac{\mu^2-3}{\mu}
	\label{existence1}
\end{equation}
or
\begin{equation}
	\mu\geq \sqrt6 \quad \wedge \quad
	\beta\leq\frac{\mu-\sqrt{\mu^2-6}}{2},
	\label{existence2}
\end{equation}
depicted graphically in Fig.~\ref{Fig-ParameterSpace-CP}, where the 
region of accelerated expansion is also shown.

\section{Perturbed scaling solutions\label{Sec-perturbed}}

Let us now analyze the fate of the scaling solutions
described in the previous section if, in Eq.~(\ref{Q_Cphi}),
instead of $C(\phi)=1$,
we allow function $C(\phi)$ to have small fluctuations around the 
constant value $1$,
namely, we choose
\begin{equation}
C(\phi) = 1 + \epsilon \sin (\alpha \kappa \phi),
\label{C perturbed}
\end{equation}
where $\epsilon$ and $\alpha$ are constants.

Contrarily to the case $C(\phi)=1$ described in previous section,
this interaction term cannot be expressed only in terms of
the variables $(x,y)$ given by Eq.~(\ref{xy});
to close the dynamical system, one needs an extra dimensionless variable $z$.
There is no standard choice for its dependence on the variables $\phi$,
$\dot{\phi}$, and $H$.
Each case is different, and $z$ must be chosen in such a way as
to make the analysis of the dynamical system feasible, as simple as
possible, and clear from the physical point of view.
To achieve this desideratum, we choose
\begin{equation}
  z = \arctan{(\kappa\phi)}, \qquad \kappa\phi = \tan{z}.
    \label{z}
\end{equation}

In the variables $x$, $y$, $z$, and $\tau$, given by Eqs.~(\ref{xy}),
(\ref{tau-t}), and (\ref{z}), the system of 
equations~(\ref{ddotphi 1})--(\ref{Friedmann 1}) gives rise now
to the three-dimensional dynamical system
\begin{subequations}
 \label{DSpss}
\begin{align}
 x_\tau = & -3x + \frac{\sqrt{6}}{2} \mu y^2 + \frac32 x(1+x^2-y^2) \nonumber
\\
          & {} + \frac{\sqrt{6}}{2} \beta 
 	         (1 - x^2 - y^2)\left[1 + \epsilon\sin (\alpha \tan z)\right],
	\label{DSpss-x}
\\
 y_\tau = & -\frac{\sqrt{6}}{2} \mu x y + \frac32 y(1+x^2-y^2),
    \label{DSpss-y}
\\
 z_\tau = & \sqrt{6}\, x(\cos z)^2.
 \label{DSpss-z}
\end{align}
\end{subequations}

For $\epsilon=0$, the evolution of $x$ and $y$ decouple from $z$.
In this case, as we have seen above,
there are two critical points corresponding to scaling solutions.
For each of these critical points, we can try to construct solutions for
nonzero $\epsilon$, by solving the evolution equations~(\ref{DSpss})
perturbatively in $\epsilon$. That is, we write
\begin{subequations}
 \label{xy-pert}
\begin{align}
 x(\tau) & = x_0 + \epsilon x_1(\tau) + \mathcal{O}(\epsilon^2),
 \label{xy-pert-x}
\\
 y(\tau) & = y_0 + \epsilon y_1(\tau) + \mathcal{O}(\epsilon^2),
 \label{xy-pert-y}
\end{align}
\end{subequations}
where $x_0$ and $y_0$ are the coordinates of the critical points at
zeroth order in $\epsilon$ given by Eqs.~(\ref{CP-A}) and (\ref{CP-B}).
At first order in $\epsilon$, Eq.~(\ref{DSpss-x}) only involves $z(\tau)$
at zeroth order in $\epsilon$. It is easy to verify from 
Eqs.~(\ref{DSpss-z}) and (\ref{xy-pert-x}) that
\begin{equation}
	\tan z(\tau) =\sqrt6 x_0 (\tau-\tau_0) + \mathcal{O}(\epsilon),
\end{equation}
where $\tau_0$ is an integration constant which, for simplicity,
we will set to zero in what follows.

Since we are interested in scaling solutions that correspond to a final 
state of accelerated expansion, let us focus our attention on the
critical point given by Eq.~(\ref{CP-B}).
At order $\epsilon$, Eqs.~(\ref{DSpss-x}) and (\ref{DSpss-y})
can be expressed as
\begin{equation}
	\frac{d}{d\tau}
	\begin{pmatrix} x_1\\y_1 \end{pmatrix} =
	J \begin{pmatrix} x_1\\y_1 \end{pmatrix} +
	\begin{pmatrix} D 
	\sin\left(\gamma\tau\right) \\0 \end{pmatrix},
	\label{first-order}
\end{equation}
with 
\begin{equation}
	D=-\frac{\sqrt6\beta(3+\beta\mu-\mu^2)}{2(\beta-\mu)^2},
	\qquad 
	\gamma=\frac{3\alpha}{\mu-\beta},
\end{equation}
and
\begin{widetext}
\begin{equation}
	J=\frac{1}{2(\mu-\beta)^2}
	\begin{pmatrix} 3(3+\beta\mu-\mu^2) & 
	\left( 2\beta^2-4\beta\mu+2\mu^2-3 \right)\sqrt{9+6\beta^2-6\beta\mu}
	\\
	\left(3+\beta\mu-\mu^2\right)\sqrt{9+6\beta^2-6\beta\mu}
	 & -3(3+2\beta^2-2\beta\mu) \end{pmatrix}.
	\label{Jacobian}
\end{equation}
\end{widetext}

The eigenvalues of $J$ are given by
\begin{equation}
 	\lambda_{\pm}=\frac{3(\mu-2\beta)}{4(\mu-\beta)}
 	\left( -1\pm\sqrt{1+F} \right), \label{eigenvalues}
\end{equation}
where
\begin{equation}
F=\frac{8(3+\beta\mu-\mu^2)
 	(3-2\beta\mu+2\beta^2)}{3(2\beta-\mu)^2}.
\end{equation}
Inside the existence region given by Eqs.~(\ref{existence1}) and
(\ref{existence2}),
the pre-factor in Eq.~(\ref{eigenvalues}) is always positive and the
expression inside the square root 
is always smaller than one,
implying that the eigenvalues are real and negative (so that 
the critical point is an attracting node) or are complex with negative real
part (i.e., the critical point is an attracting spiral).
At the boundary of the existence region,
given by Eqs.~(\ref{existence1}) and (\ref{existence2}),
the stability cannot be decided at first order;
one must go beyond linear theory (for a detailed analysis 
of the stability at the boundary using center manifold theory see
Ref.~\cite{sa-2021}).

The general solution to Eq.~(\ref{first-order}) is given by the sum 
of a particular solution and a linear combination of the solutions
to the homogeneous equation. The latter will converge to the critical point
if the parameters $\mu$ and $\beta$ lie inside the existence region given
by Eqs.~(\ref{existence1}) and (\ref{existence2}) and amount
to a linear combination of two eigenvectors multiplied by exponentials
of the corresponding eigenvalues times $\tau$. We can take the following
ansatz for a particular solution,
\begin{align}
	x_1(\tau)=a\sin(\gamma\tau)+b\cos(\gamma\tau),
	\\
	y_1(\tau)=c\sin(\gamma\tau)+d\cos(\gamma\tau),
\end{align}
which represents a periodic solution oscillating around the origin with
the same frequency as the inhomogeneous term (i.e., the coupling to $z$
makes $x$ and $y$ fluctuate around the original critical point).
Substituting into Eq.~(\ref{first-order}) yields the constants
$a$, $b$, $c$, and $d$ in terms of the coefficients of the matrix $J$
given by Eq.~(\ref{Jacobian}); this can be evaluated explicitly in terms
of $\mu$ and $\beta$, but the expressions are lengthy and not illuminating.

From this analysis we conclude that if the interaction term $Q$ is 
slightly perturbed (see Eq.~(\ref{C perturbed})),
the dynamical system still admits scaling critical points, in particular,
the one corresponding to late-time accelerated expansion.
This circumstance is worth emphasizing, because, as we will see in
Sec.~\ref{Sec-DSsolutions},
if the function $C(\phi)$ in the interaction term is chosen
to be unbounded, as, for instance, in the case $C(\phi)=(\kappa \phi)^n$,
the critical points corresponding to scaling solutions disappear
altogether.

%%%%%%% DISPLACED TABLE %%%%%%%%%

\begin{table*}[t]
\begin{tabular}{ccccccccc}
\hline\hline
Line/point
  & \quad $x$
  & \quad $y$
  & \quad $z$
  & \quad Existence
  & \quad $\Omega_\phi$
  & \quad $\Omega_\texttt{DM}$
  & \quad $w_{\rm eff}$
  & \quad Acceleration \\ \hline \noalign{\vskip 1mm}
$A$ & \quad $x$
  & \quad $\sqrt{1-x^2}$
  & \quad $-\pi/2$
  & \quad $|x|\leq1$, $\mu>0$, $\beta\neq0$
  & \quad $1$
  & \quad $0$
  & \quad $2x^2-1$
  & \quad $|x|<1/\sqrt3$ \\ \noalign{\vskip 1mm}
$B$ & \quad $x$
  & \quad $\sqrt{1-x^2}$
  & \quad $\pi/2$
  & \quad $|x|\leq1$, $\mu>0$, $\beta\neq0$
  & \quad $1$
  & \quad $0$
  & \quad $2x^2-1$
  & \quad $|x|<1/\sqrt3$ \\ \noalign{\vskip 1mm}
$C$ & \quad $0$
  & \quad $0$
  & \quad $0$
  & \quad $\mu>0$, $\beta\neq0$
  & \quad $0$
  & \quad $1$
  & \quad $0$
  & \quad never \\ \noalign{\vskip 1mm}
    \hline\hline
\end{tabular}
 \caption{\label{Table:properties CP} Properties of the critical
lines/points of the dynamical system~(\ref{DSn}).}
\end{table*}

\section{Coupled quintessence with a generalized interaction term\label{Sec-CQgeneralized}}

Let us consider the interaction term between the dark-energy scalar
field and the dark-matter fluid to be of the form
given by Eq.~(\ref{Q_Cphi}) with\footnote{
A different choice of $C(\phi)$, namely,
$C(\phi)\propto \bigl(V(\phi)\bigr)^n$, was made 
in Ref.~\cite{lopez-2010}.}
\begin{equation}
  C(\phi)=(\kappa \phi)^n, \qquad n=1,2,3,\ldots.
  \label{Cphi}
\end{equation}
Here, as in the case considered in the previous section,
the interaction term $Q$ cannot be expressed
only in terms of the variables $(x,y)$ given by Eq.~(\ref{xy}), requiring
the introduction of a new dimensionless variable $z$, which we choose, 
again, to be given by Eq.~(\ref{z}).
This choice allows for a compactification of the phase space; values of
$\phi$ between $-\infty$ and $+\infty$ correspond to values of $z$ between
$-\pi/2$ and $\pi/2$.
We choose to augment the phase space with the boundaries $z=\pm\pi/2$. In
these boundaries, the orbits of the dynamical system are not physical, but,
by continuity arguments, they allow for the determination of the behavior
of the physical ones, lying in the region $|z|<\pi/2$.
With this assumption, the phase space of the dynamical system is then the 
half-cylinder $\{ (x,y,z)| x^2+y^2\leq1, y\geq0,-\pi/2\leq z\leq\pi/2\}$.

To avoid singularities in the dynamical system, we also introduce
a new time variable,
\begin{equation}
  \qquad \frac{d\eta}{dt}=\frac{H}{(\cos{z})^n}.
  \label{eta-t}
\end{equation}
Note that in the case $C(\phi)=1$ 
discussed in Sec.~\ref{Sec-standardCQ}, the coordinate $\tau$ had
a simple physical interpretation (see Eq.~(\ref{tau-t}));
$\tau = \ln a$ was simply the so-called e-fold number,
a convenient measure of expansion.
Here, because of the extra term $(\cos{z})^n$,
the coordinate $\eta$ has not such a simple physical
interpretation.

In the variables $x$, $y$, $z$, and $\eta$, the
system of equations~(\ref{ddotphi 1})--(\ref{Friedmann 1})
gives rise to the three-dimensional dynamical system
\begin{subequations}
 \label{DSn}
\begin{align}
 x_\eta = & (\cos{z})^n \left[
      -3x + \frac{\sqrt{6}}{2} \mu y^2 + \frac32 x(1+x^2-y^2) \right]
      \nonumber
\\ 
 & +\frac{\sqrt{6}}{2} \beta (1-x^2-y^2)(\sin{z})^n, \label{DSn-x}
\\
 y_\eta = &(\cos{z})^n \left[
      -\frac{\sqrt{6}}{2} \mu x y + \frac32 y(1+x^2-y^2) \right],
      \label{DSn-y}
\\
 z_\eta = & \sqrt{6}\, x(\cos{z})^{n+2}
 \label{DSn-z},
\end{align}
\end{subequations}
where the subscript $\eta$ denotes a derivative with respect to this variable.

The above dynamical system contains two dimensionless constants, $\mu$ and
$\beta$.
Without any loss of generality, we can assume $\mu$ to be non-negative,
just like for the dynamical system~(\ref{DS0}).
Indeed, the dynamical system is invariant under the transformation
$x\rightarrow-x$, $z\rightarrow-z$, $\mu\rightarrow-\mu$, and
$\beta\rightarrow (-1)^{n+1}\beta$;
this implies that
solutions for negative values of $\mu$ can be obtained straightforwardly from
solutions for positive values of $\mu$, provided reflections over $x$ and $z$
are performed (if only even values of $n$ were considered, we could choose
$\beta$ to be non-negative instead of $\mu$). Furthermore, since we want to
analyze cosmological models with a direct interaction between dark matter and
dark energy and also with a non-constant potential, we impose $\beta\neq0$ and
$\mu\neq0$. Therefore, the parameter space of our model is the region
$\{(\mu,\beta)| \mu>0, \beta\neq0\}$.

In terms of the variables $x$ and $y$ the density parameters
for dark matter
and dark energy, $\Omega_\texttt{DM}$ and $\Omega_\phi$, and the effective
equation-of-state parameter, $w_{\rm eff}$, are given by
\begin{equation}
 \Omega_\texttt{DM}=1-x^2-y^2,
    \label{Omegaxi}
\end{equation}
\begin{equation}
 \Omega_\phi=1-\Omega_\texttt{DM}=x^2+y^2,
    \label{Omegaphi}
\end{equation}
\begin{equation}
 w_{\rm eff}=\frac{p_\phi}{\rho_\texttt{DM}+\rho_\phi}=x^2-y^2,
    \label{weff}
\end{equation}
where
\begin{equation}
 \rho_\phi=\frac{\dot{\phi}^2}{2}+V_a e^{-\mu\kappa\phi}
\quad {\rm and} \quad
 p_\phi=\frac{\dot{\phi}^2}{2}-V_a e^{-\mu\kappa\phi}
\end{equation}
are, respectively, the energy density and the pressure of the scalar
field $\phi$. Note that the above expressions do not depend on $z$
and they coincide with those of the coupled quintessential cosmological
model given by Eqs.~(\ref{DS0}).

%%%%%%%%%%%%% DISPLACED FIGURES %%%%%%%%%%%%%%

\begin{figure*}[t]
 \includegraphics{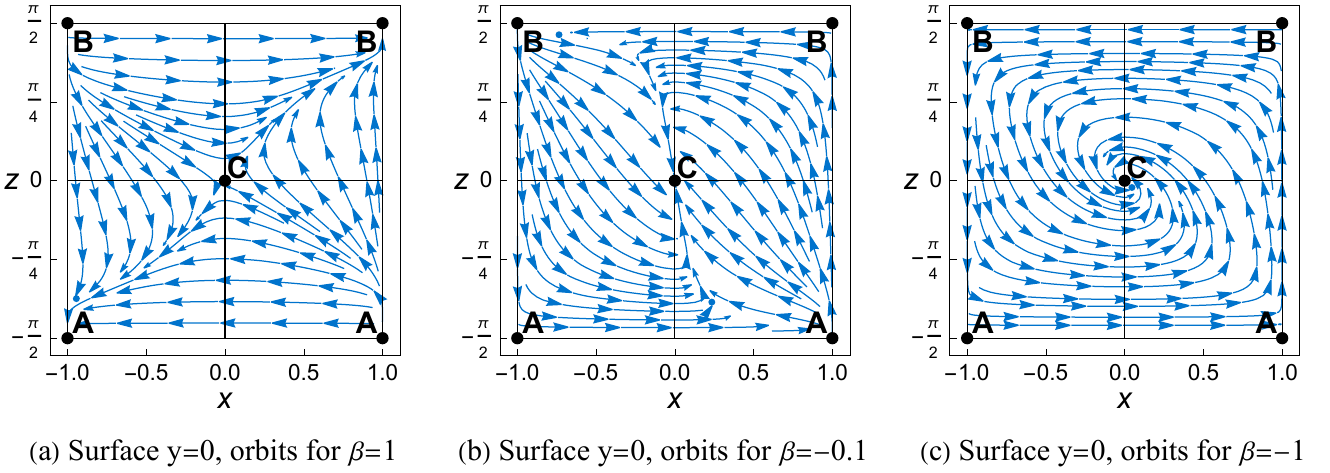}
 \caption{\label{fig-n1-back} Orbits of the dynamical system~(\ref{DSn})
on the surface $y=0$, for $n=1$ and $\beta=1$ [panel~(a)], $\beta=-0.1$
[panel~(b)], $\beta=-1$ [panel~(c)] (on this surface, the orbits do not 
depend on the parameter $\mu$). The critical lines/points $A$, $B$, and $C$
are represented by black dots.}
\end{figure*}

\begin{figure*}[t]
 \includegraphics{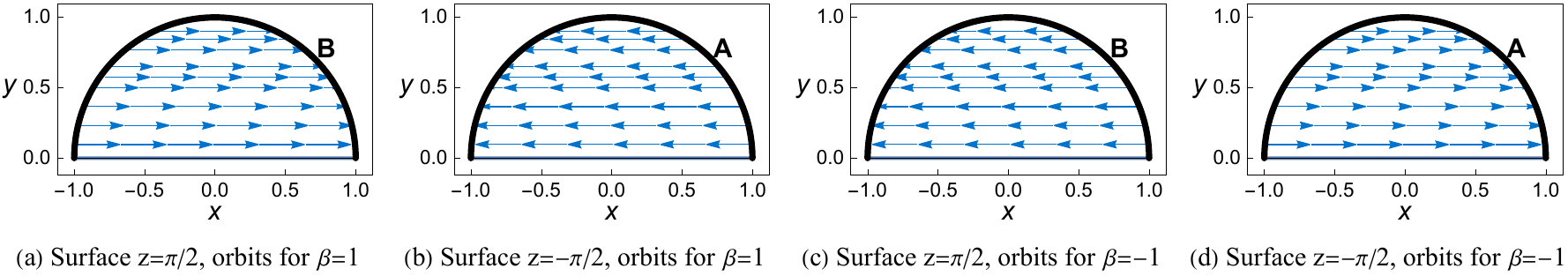}
 \caption{\label{fig-n1-top-bottom} Orbits of the dynamical
system~(\ref{DSn}) on the surfaces $z=\pm\pi/2$, for $n=1$ and
$\beta=1$ [panels~(a) and (b)], $\beta=-1$ [panels~(c) and (d)]
(on this surface, the orbits do not depend on the parameter $\mu$).
The critical lines $A$ and $B$ are represented by thick black lines. }
\end{figure*}

\begin{figure*}[t]
 \includegraphics{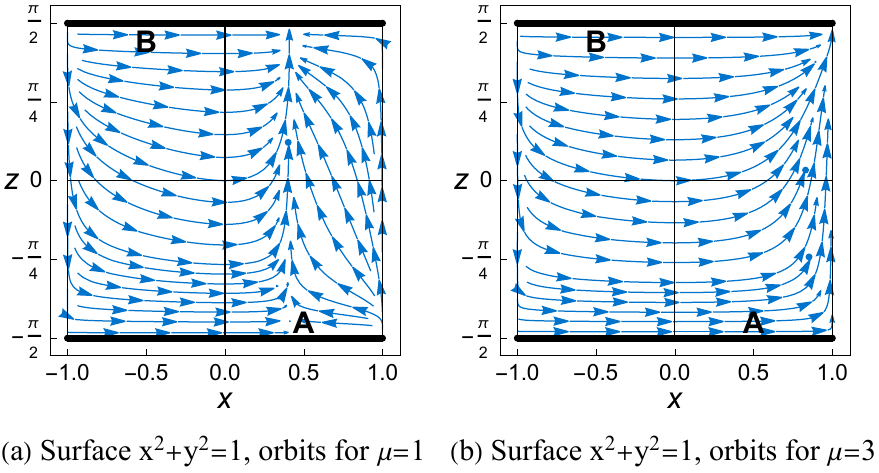}
 \caption{\label{fig-n1-front} Orbits of the dynamical system~(\ref{DSn}) on the
surface $x^2+y^2=1$, for $n=1$ and $\mu=1$ [panel~(a)], $\mu=3$ [panel~(b)]
(on this surface, the orbits do not depend on the parameter $\beta$). Critical
lines $A$ and $B$ are represented by thick black lines.}
\end{figure*}

\section{The dynamical system solutions\label{Sec-DSsolutions}}

The system of differential equations~(\ref{DSn}) can be investigated 
using the powerful methods of qualitative analysis of dynamical systems
(for a recent review on the application of such methods to cosmology and,
in particular, to coupled quintessential models, see 
Ref.~\cite{bahamonde-2018}).

The dynamical system~(\ref{DSn}) has two critical lines (continuous collections
of critical points), $A(x,\sqrt{1-x^2},-\pi/2)$ and
$B(x,\sqrt{1-x^2},\pi/2)$, as well as a critical point, $C(0,0,0)$.

The properties of these critical lines/points are listed on 
Table~\ref{Table:properties CP}. They exist for all values of
$\mu$ and $\beta$ belonging to the parameter space
($\mu>0$ and $\beta\neq0$).
Note that in the vicinity of the critical point $C$, the solution is
matter dominated, behaving like dust. On the contrary, near critical lines $A$
and $B$, the solution is dominated by the scalar field $\phi$, which,
depending on the specific points considered on these lines, can have any
behavior ranging from stiff matter ($x=\pm1$) to dark energy ($|x|<1/\sqrt3$).
An important conclusion we can draw is that none of the critical
points/lines of the dynamical system~(\ref{DSn}) correspond to scaling solutions.

The stability properties of the critical lines/points are investigated
in the Appendix~\ref{Sec-appendix}, where we specialize for the case $n=1$
(note, however, that the analysis can be straightforwardly adapted
to the cases $n=2,3,4,\dots$).
Because the linear theory does not suffice to investigate the stability
properties of the critical lines $A$ and $B$, we have to resort to other
methods, in particular, to the center manifold theory.

The stability analysis performed in the  Appendix~\ref{Sec-appendix}
reveals that, for $\beta>0$, 
the dynamical system~(\ref{DSn}) has just one attracting critical
point: $B(\mu/\sqrt6,\sqrt{1-\mu^2/6},\pi/2)$ for $0<\mu<\sqrt6$ or
$B(1,0,\pi/2)$ for $\mu\geq\sqrt6$.
All other critical points are unstable (either repeller or saddle points);
in particular, for $0<\mu<\sqrt6$ there are two repellers, namely, 
$A(1,0,-\pi/2)$ and $B(-1,0,\pi/2)$, and for $\mu\geq\sqrt6$ there is just 
one repeller, namely, $B(-1,0,\pi/2)$. 
All orbits converge asymptotically to the global attractor,
with the exception of the heteroclinic ones, connecting critical points
along the boundaries of the phase space.

\begin{figure*}[t]
 \includegraphics{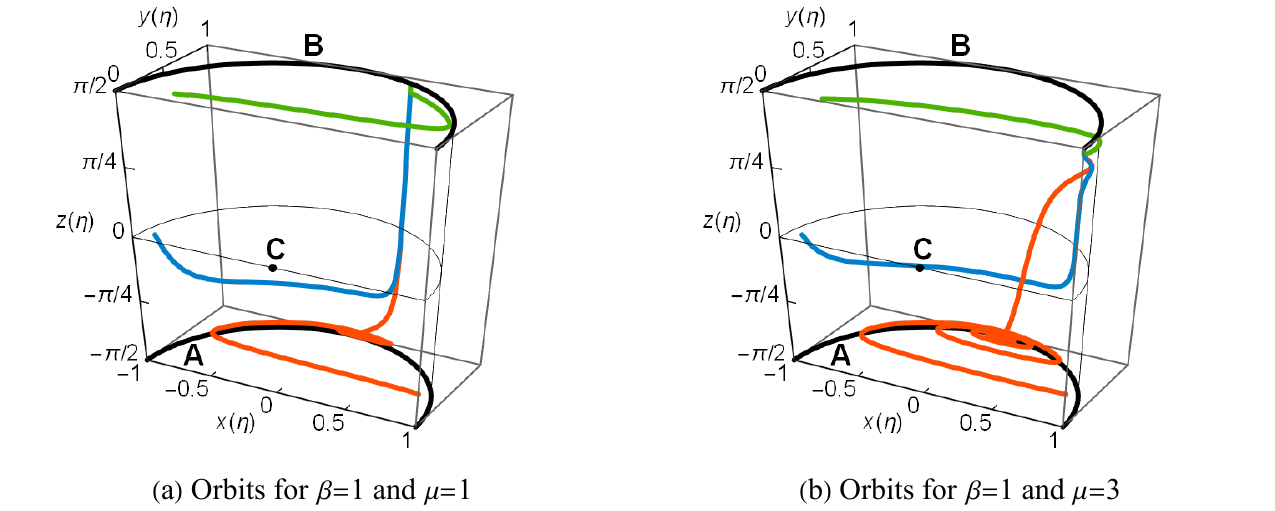}
\caption{\label{fig-full-n1-beta+} For $\beta>0$, the dynamical 
system~(\ref{DSn}) has just one attracting critical point,
$B(\mu/\sqrt6,\sqrt{1-\mu^2/6},\pi/2)$ for $0<\mu<\sqrt6$ 
or $B(1,0,\pi/2)$ for $\mu\geq\sqrt6$.
Some orbits are shown on the phase space for $\mu=1$ (left panel) 
and $\mu=3$ (right panel). In both panels $\beta=1$.
The critical lines $A$ and $B$ are represented by thick black lines
and the critical point $C$ by a black dot.}
\end{figure*}

\begin{figure*}[t]
  \includegraphics{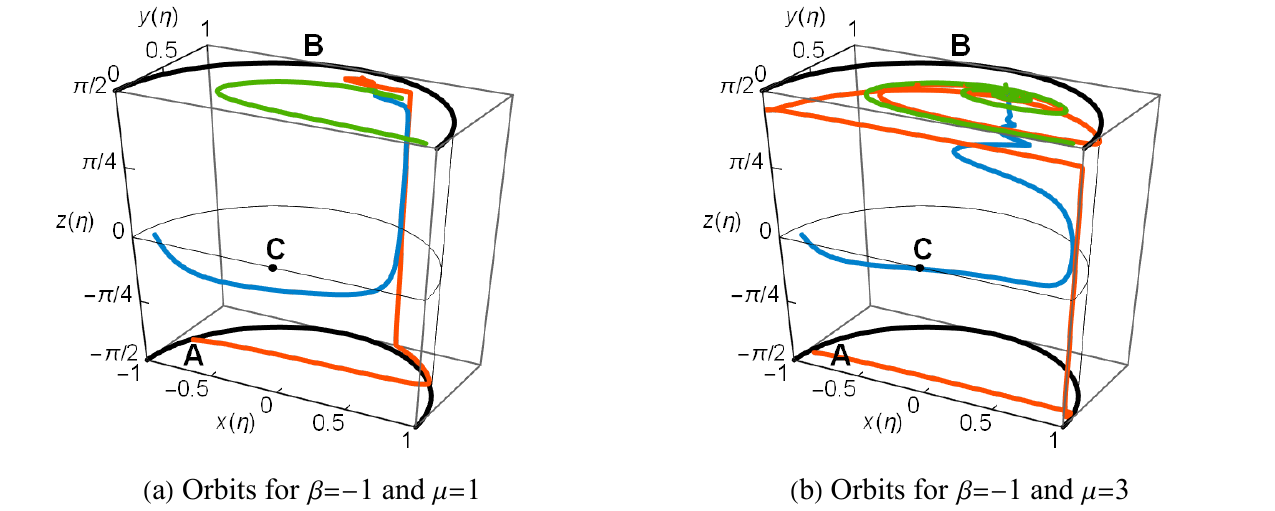}   
\caption{\label{fig-full-n1-beta-} For $\beta<0$, the dynamical 
system~(\ref{DSn}) has just one attracting critical point,
$B(0,1,\pi/2)$. 
Some orbits are shown on the phase space for $\mu=1$ (left panel) 
and $\mu=3$ (right panel). In both panels $\beta=-1$.
The critical lines $A$ and $B$ are represented by thick black lines
and the critical point $C$ by a black dot.}
\end{figure*}

For $\beta<0$, the dynamical system~(\ref{DSn}) has also just one
attractor: the critical point $B(0,1,\pi/2)$; all orbits, but the
heteroclinic, converge to this point.
However, this convergence proceeds very slowly; as shown in the
Appendix~\ref{Sec-appendix},
orbits first quickly approach a point with coordinates
$x$ and $y$ given Eqs.~(\ref{uvw}), (\ref{u0}), and (\ref{v0}) and 
$z\lesssim\pi/2$, after which a long time is then required for them 
to converge to the critical point $B(0,1,\pi/2)$.

Note that accelerated expansion occurs whenever orbits
enter the region of the phase space defined by 
$\{(x,y,z)|x^2-y^2<-1/3; -\pi/2\leq z\leq\pi/2 \}$. Of particular
relevance is the fact that for $\beta>0$ the global attractor	
$B(\mu/\sqrt6,\sqrt{1-\mu^2/6},\pi/2)$ lies inside that region for 
$\mu<\sqrt2$, while for $\beta<0$ the global attractor $B(0,1,\pi/2)$
is always inside that region for any allowed value of $\mu$.

Orbits lying on the two-dimensional surfaces $y=0$, $x^2+y^2=1$, and 
$z=\pm\pi/2$, which delimit the phase space, are shown in
Figs.~\ref{fig-n1-back}--\ref{fig-n1-front} for $n=1$ and relevant values of
the parameters $\mu$ and $\beta$.
Some other orbits, lying inside the phase space,
are shown in Figs.~\ref{fig-full-n1-beta+} and Fig.~\ref{fig-full-n1-beta-}.

\section{Solutions of cosmological relevance \label{Sec-cosmological}}

To be of cosmological relevance, solutions of the dynamical 
system~(\ref{DSn}) should contain, at later times, a long enough 
matter-dominated period followed by an era of accelerated expansion.

According to current cosmological data, the matter-dominated
period is comprised in the interval $0.5\lesssim\zeta\lesssim3000$, 
where $\zeta=a_0/a-1$ is the redshift.
Since this quantity is related to the time variable $\eta$ by
(see Eq.~(\ref{eta-t}))
\begin{equation}
	\ln(\zeta+1)= \ln \left( \frac{a_0}{a} \right)
	            = \int_\eta^0 [\cos z(\eta)]^n d\eta,
	\label{redshift}     
\end{equation}
cosmologically relevant solutions should have a matter-dominated
period starting at $\eta_i$, such that 
$\int_{\eta_i}^0 [\cos z(\eta)]^n d\eta\approx8$, and ending at
$\eta_f$, such that 
$\int_{\eta_f}^0 [\cos z(\eta)]]^n d\eta\approx0.4$.
These conditions can be satisfied only for orbits of the phase space that
pass very closely to the critical point $C$ and, consequently, stay long
enough near this point. There, $z\approx0$, implying that 
$\int_\eta^0 [\cos z(\eta)]^n d\eta \approx \int_\eta^0 d\eta=-\eta$, or
$\eta_i\approx-8$ and $\eta_f\approx-0.4$.
In all our numerical simulations, presented below, the matter-dominated
era starts at $\eta_i\approx-8$ and the transition from this era to an 
era of accelerated expansion occurs at $\eta_f\approx-0.4$.

Furthermore, a current era of accelerated expansion exists if the 
$x$-coordinate of the global attractor satisfies the condition
$|x|<1/\sqrt3$. For $\beta>0$ this requires $\mu<\sqrt2$, while
for $\beta<0$ this condition is satisfied for any value of $\mu$.

On top of the above requirements, recent cosmological observations also
require the density parameters for dark energy and dark matter 
at the present time to be 
$\Omega_{\phi}(\eta=0)\approx0.69$ and 
$\Omega_\texttt{DM}(\eta=0)\approx0.31$.
Note that in our coupled quintessence model, for simplicity, 
we have not introduced an explicit baryonic matter component, instead we 
consider it as a part --- albeit small --- of $\rho_\texttt{DM}$.

Let us present some explicit representative examples of solutions 
that satisfy the above requirements and, therefore,
are of cosmological relevance.
Here, again, we specialize to the case $n=1$.

\begin{figure}[t]
 \includegraphics{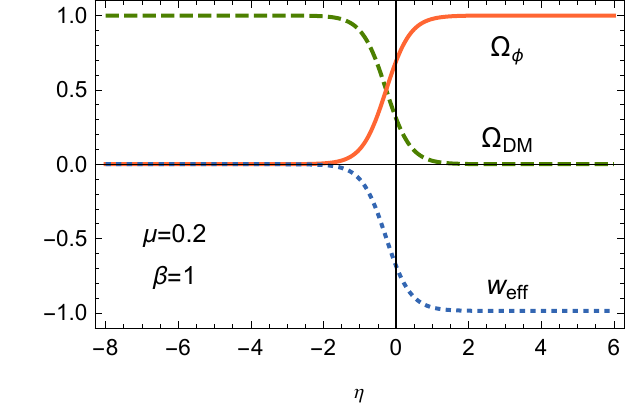}
\caption{\label{fig-cosmo11}
Evolution of the density parameters $\Omega_\phi$
and $\Omega_\texttt{DM}$ and the effective equation-of-state parameter
$w_{\rm eff}$ for $\beta=1$ and $\mu=0.2$, showing a long enough
matter-dominated era followed by an era of everlasting accelerated expansion.
At the present time ($\eta=0$), $\Omega_\phi\approx0.69$,
$\Omega_\texttt{DM}\approx0.31$, and $w_{\rm eff}\approx-0.68$.
In the future, $\Omega_\phi$, $\Omega_\texttt{DM}$, and $w_{\rm eff}$
quickly approach 1, 0, and $-0.99$, respectively.}
\end{figure}

First, we consider the case $\beta>0$ and $0<\mu<\sqrt2$.
Choosing $\beta=1$, $\mu=0.2$, $x_i=y_i=9.23\times10^{-6}$, and $z_i=0$,
we obtain for the density parameters $\Omega_\phi$
and $\Omega_\texttt{DM}$ and for the effective equation-of-state parameter
$w_{\rm eff}$ the evolution depicted in 
Fig.~\ref{fig-cosmo11}. Note that the initial values for $x$,
$y$ and $z$ must be taken sufficiently
close to the critical point $C$
in order to guarantee a long enough matter-dominated era,
starting at $\eta\approx-8$ and ending at $\eta\approx-0.4$.
Furthermore, adopting the convention that the present
time corresponds to $\eta=0$, specific choices of the initial values for
$x$, $y$ and $z$ are required to guarantee
$\Omega_\phi(\eta=0)\approx0.69$ and $\Omega_\texttt{DM}(\eta=0)\approx0.31$.
This does not represent fine tuning;
indeed, we could choose some other initial values 
for this variables (provided they are close enough to the critical point $C$)
and identify the present time with the value of
$\eta$ for which the energy density for dark energy is $69\%$
of the total energy density. 

In the above example, in the limit $\eta\to\infty$, $\Omega_\phi$ and
$\Omega_\texttt{DM}$ quickly approach 1 and 0, respectively,
while $w_{\rm eff}$ tends to $-0.99$; this corresponds to accelerated
expansion in a Universe completely dominated by dark energy.

\begin{figure}[t]
 \includegraphics{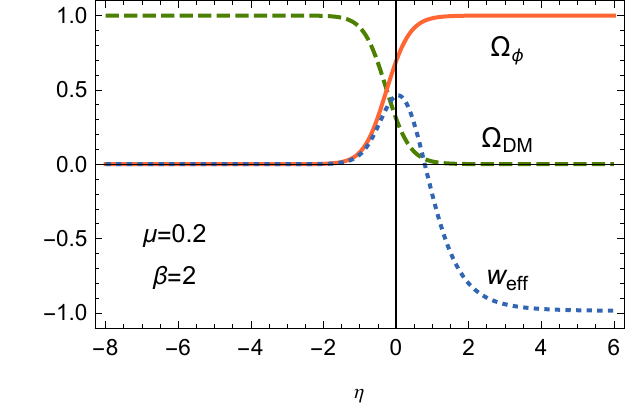}
\caption{\label{fig-cosmo12}
Evolution of the density parameters $\Omega_\phi$
and $\Omega_\texttt{DM}$ and the effective equation-of-state parameter
$w_{\rm eff}$ for $\beta=2$ and $\mu=0.2$.
Between the matter-dominated era and the accelerated-expansion era,
there is a period of dominance of the kinetic term of the scalar field
$\phi$ (kination), for which $w_{\rm eff}$ takes positive values.}
\end{figure}

Note that the asymptotic value of $w_{\rm eff}$ depends only on the 
parameter $\mu$;
indeed, for the critical point $B(\mu/\sqrt6,\sqrt{1-\mu^2/6},\pi/2)$,
equation~(\ref{weff}) yields
\begin{equation}
	w_{\rm eff}=-1+\frac{\mu^2}{3}.
\end{equation}
On the other hand, the behavior of $w_{\rm eff}$ during the present era
is determined by the parameter $\beta$.
Indeed, for higher values of $\beta$, the transition from matter
domination to accelerated expansion driven by dark energy does not
proceed directly; at the present era, the Universe undergoes an
intermediate stage of dominance by the kinetic term of the scalar
field $\phi$ (kination).
As an example of such behavior consider the solution obtained
for $\beta=2$, $\mu=0.2$, $x_i=y_i=2.5\times10^{-6}$, and $z_i=0$,
for which $\Omega_\phi(\eta=0)\approx0.69$,
$\Omega_\texttt{DM}(\eta=0)\approx0.31$, and 
$w_{\rm eff}(\eta=0)\approx0.46$ (see Fig.~\ref{fig-cosmo12}).
In order to avoid this kination period, the value of the parameter
$\beta$ cannot be too high; our numerical simulation show that
$w_{\rm eff}(\eta=0)<-1/3$ for $\beta\lesssim1.7$.

Let us now consider the case $\beta<0$.
For $\beta=-1$, $\mu=1$, $x_i=y_i=1.05\times10^{-5}$, and $z_i=0$ we obtain
a solution with an adequate duration of the matter-dominated era and also
with $\Omega_\phi(\eta=0)\approx0.69$, $\Omega_\texttt{DM}(\eta=0)\approx0.31$,
and $w_{\rm eff}(\eta=0)\approx-0.61$
(see Fig.~\ref{fig-cosmo-11}).
As shown in the Appendix,
the global attractor is now the critical point $B(0,1,\pi/2)$, but the 
orbits do not converge immediately to this point.
They first rapidly approach a temporary state,
from where they advance very slowly to the final critical point.
In order to clarify what {\textit{very slowly} means,
let us point out that,
for an increase of $\eta$ from 20 to 66, corresponding to a growth of
the scale factor by $10^3$, the $x$-variable decreases from $0.20$
to just $0.14$; even for a future increase of the scale factor by $10^{50}$,
corresponding to $\Delta \eta\approx2000$, $x$
decreases just by three quarters ($x(\eta_f)\approx0.05$).

Let us emphasize that, during the temporary state, the above solution
mimics the behavior of an accelerated scaling solution 
for which $\Omega_\texttt{DM}/\Omega_\phi\sim\mathcal{O}(1)$.
As our analysis of the dynamical system~(\ref{DSn}) has shown,
this temporary state is not an attractor,
but the approach from there to the final critical point lasts so long,
that, from a physical perspective, it could be considered as such.

To conclude, we point out that, in the case $\beta<0$, 
the allowed values of the parameter $\mu$
are constrained by the requirement that, at the present time, 
$\Omega_\phi$ should represent approximately $69\%$ of the total
energy density of the Universe.
To see this, consider the case $\beta=-1$ and $\mu=3$, for which,
no matter the initial values of $x_i$, $y_i$, and $z_i$, the
density parameter for dark energy at present time is substantially
smaller than the value required from observations
(see Fig.~\ref{fig-cosmo-13}).
To avoid this problem, $\mu$ should be smaller than about $1.9$.

\begin{figure}[t]
 \includegraphics{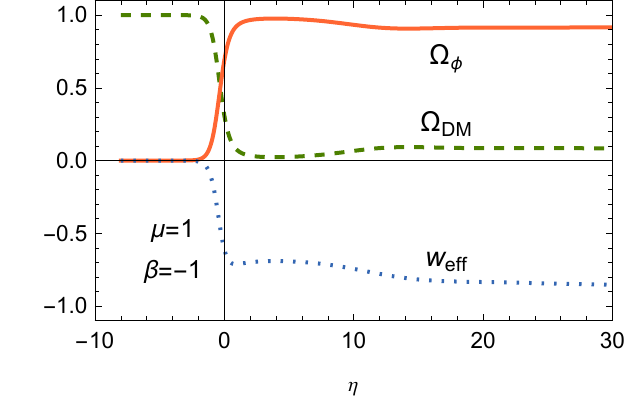}
\caption{\label{fig-cosmo-11} 
Evolution of the density parameters $\Omega_\phi$
and $\Omega_\texttt{DM}$ and the effective equation-of-state parameter
$w_{\rm eff}$ for $\beta=-1$ and $\mu=1$.
In its approach to the final state, for which
$\Omega_\phi=1$ and $\Omega_\texttt{DM}=0$, the Universe
stays for a long time in a temporary state that corresponds to an
accelerated scaling solution.}
\end{figure}

\begin{figure}[t]
 \includegraphics{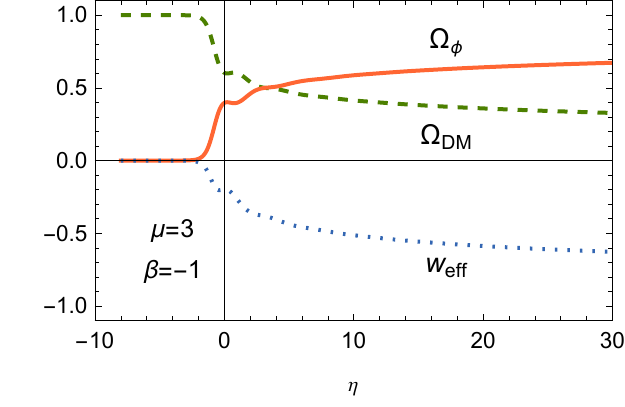}
\caption{\label{fig-cosmo-13} Evolution of the density parameters
$\Omega_\phi$ and $\Omega_\texttt{DM}$ and the effective equation-of-state
parameter $w_{\rm eff}$ for $\beta=-1$, $\mu=3$,
$x_i=y_i=1.36\times10^{-5}$, and $z_i=0$.
The density parameter for dark energy at present time is substantially
smaller than the value required from observations. }
\end{figure}

In this section, we have shown that the dynamical system~(\ref{DSn}) admits
solutions that can reproduce, at least qualitatively, the late-time 
evolution of the Universe, namely, they give rise, successively,
to a long enough matter-dominated era and an era of accelerated expansion.

\section{Conclusions \label{Sec-conclusions}}

In this work we studied a class of cosmological models with a scalar
field representing quintessence coupled to dark matter,
which is treated as a perfect fluid.
Taking the interaction term between dark energy and dark 
matter to be $Q \propto\rho_\texttt{DM} C(\phi) \dot{\phi}$,
we have identified and analyzed solutions of cosmological relevance.

We showed that, for small perturbations of the coupling term,
namely, $C(\phi) = 1 + \epsilon \sin (\alpha \kappa \phi)$,
scaling solutions with late-time accelerated expansion continue to exist,
even though they acquire small modifications.
However, if $C(\phi)$ is an unbounded function 
such as $(\kappa \phi)^n$, $n=1,2,3,\ldots$,
it turns out that those scaling solutions disappear altogether.

A detailed dynamical-system analysis was done for the case
$C(\phi) = \kappa \phi$,
using the dynamical variables $\{x,y,z\}$ defined by Eqs.~(\ref{xy})
and (\ref{z}), and the transformed time variable $\eta$ defined by 
Eq.~(\ref{eta-t}).
Our conclusion is that in each case there is one attracting critical point,
determining the late-time evolution of the universe.
For $\beta>0$ it is at $(x,y,z)=(\mu/\sqrt{6},\sqrt{1-\mu^2/6},\pi/2)$
for $0<\mu<\sqrt{6}$, and at $(x,y,z)=(1,0,\pi/2)$ for $\mu\ge\sqrt{6}$.
On the other hand, for $\beta<0$, all orbits converge to 
$(x,y,z)=(0,1,\pi/2)$.
All other critical points, isolated or non-isolated, are
repellers or saddle points.
Note that attractors with $|x|<1/\sqrt{3}$ involve solutions with 
late-time accelerated expansion.

A careful analysis showed also that the dynamical system admits solutions
that are cosmologically relevant, at least qualitatively,
in that, at later times, they give rise to a long enough matter-dominated era,
followed by an era of accelerated expansion, 
both for the cases $\beta>0$ and $\beta<0$.
A surprising phenomenon was observed in the latter case,
in that the approach of the orbits to the final critical point occurs in two
phases: an exponential approach at approximately constant $z = z_0 < \pi/2$
toward a point $(x_0,y_0,z_0)$ followed by an extremely slow evolution along
an approximately straight line toward the final critical point.
During this approach,
the density parameters $\Omega_\phi$ and $\Omega_{\texttt{DM}}$
take values between $0$ and $1$, and therefore, for all practical purposes,
this final stage could be considered as an accelerated scaling solution.
Thus, despite the fact that, strictly,
there are no accelerated scaling solutions in the coupled quintessential
cosmological model with $C(\phi)=(\kappa \phi)^n$,
$n=1,2,3,\ldots$, there are solutions that mimic their behavior.

The analysis in this work restricted the form of the function $C(\phi)$
to small (harmonic) fluctuations around a constant, as well as positive
integer powers of $\phi$.
A full analysis of all possible forms of the function $C(\phi)$ might be
very interesting.

\begin{acknowledgments}
R.~P.\ acknowledges financial support from Funda\c c\~ao para a Ci\^encia e
a Tecnologia (Portugal) under grant UIDB/00099/2020.
P.~M.~S.\ acknowledges support from Funda\c{c}\~ao para a Ci\^encia e a
Tecnologia (Portugal) through the research grants UIDB/04434/2020
and UIDP/04434/2020.

\end{acknowledgments}

\appendix
\section{Analysis of the dynamical system\label{Sec-appendix}}

In this Appendix, we investigate the stability properties of
the critical lines/points $A(x,\sqrt{1-x^2},-\pi/2)$,
$B(x,\sqrt{1-x^2},\pi/2)$, and $C(0,0,0)$ of the dynamical system~(\ref{DSn})
resorting to methods of qualitative analysis of dynamical systems
(see, for instance,
Refs.~\cite{carr-1982,guckenheimer-1983,bogoyavlensky-1985}, as well as a
recent review on dynamical systems applied to cosmology \cite{bahamonde-2018}).

\subsection{Critical line $A(x,\sqrt{1-x^2},-\pi/2$)\label{Sec-A}}

Let us consider a specific point $A(x_c,\sqrt{1-x_c^2},-\pi/2)$ of the
critical line $A(x,\sqrt{1-x^2},-\pi/2)$, where $x_c\in [-1,1]$ is a constant.
The Jacobian matrix of the dynamical system~(\ref{DSn}), given by
\begin{equation}
 J=\begin{pmatrix}
     \sqrt6\beta x_c & \sqrt6\beta\sqrt{1-x_c^2} &
     -3(1-x_c^2) \left( x_c-\frac{\mu}{\sqrt6} \right) \\
     0 & 0 & 3 x_c \sqrt{1-x_c^2} \left( x_c-\frac{\mu}{\sqrt6} \right) \\
     0 &0& 0
    \end{pmatrix},
   \label{jacA}
\end{equation}
has eigenvalues $\lambda_1=\sqrt6\beta x_c$, $\lambda_2=0$, and 
$\lambda_3=0$.

A non-isolated critical point for which the Jacobian matrix has just one zero
eigenvalue is called normally hyperbolic and the stability of the orbits along
the remaining directions can be determined by applying the linear stability
theory (see, for instance, Ref.~\cite{bouhmadi-2017} for an example in a
cosmological context).
Since, in our case, the Jacobian matrix has two zero eigenvalues, the linear
stability theory does not suffice and, consequently, one should resort to
alternative methods of analysis. In what follows, we will use the center
manifold theory.
Note that for $x_c=0$ all eigenvalues vanish and the center manifold theory
cannot be applied either, requiring us to use other methods to investigate
the stability properties of the critical point $A(0,1,-\pi/2)$.

The conclusion we can extract from linear stability theory is 
that, for $\beta>0$,
along the $x$-direction the orbits are attracted to the critical line 
if $-1\leq x_c<0$ or repelled if $0<x_c \leq 1$ (see panel (b) of
Fig.~\ref{fig-n1-top-bottom}).
For $\beta<0$, the behavior is the opposite 
(see panel (d) of Fig.~\ref{fig-n1-top-bottom}).
To study the behavior of the orbits for the other directions,
we resort to the center manifold theory for $x_c\neq0$
(see comment above) and to other methods for $x_c=0$.

First, we consider the case $x_c\neq 0$.

In the dynamical system~(\ref{DSn}), after expanding 
$\sin z$ and $\cos z$ in Taylor series about $z=-\pi/2$,
we introduce new variables
\begin{equation}
 u=x-x_c, \quad v=y-\sqrt{1-x_c^2}, \quad w=z+\pi/2,
 \label{cv1}
\end{equation}
which shift the critical point $A(x_c,\sqrt{1-x_c^2},-\pi/2)$ to the origin.
In these new coordinates, the dynamical system becomes
\begin{subequations}
\label{u,v,w-eta-n=1}
\begin{align}
 u_\eta & = \sqrt6\beta x_c u + \sqrt6\beta\sqrt{1-x_c^2} \, v \nonumber
\\
 & \hspace{4mm}­ -3(1-x_c^2) \left( x_c-\frac{\mu}{\sqrt6} \right)w 
 + f_1(u,v,w),
\\
 v_\eta & = 3 x_c \sqrt{1-x_c^2} \left( x_c-\frac{\mu}{\sqrt6} \right)w +
 f_2(u,v,w),
\\
 w_\eta & = f_3(u,v,w),
\end{align}
\end{subequations}
where $f_i$ are $\mathcal{O}(u^2, v^2,w^2,uv,uw,vw)$.

To bring the above dynamical system to the standard form required to apply
the center manifold theorem, we perform another change of variables, namely,
\begin{equation}
 \begin{pmatrix}
  u\\v\\w
 \end{pmatrix}
  = S
 \begin{pmatrix}
 U\\V\\W
 \end{pmatrix}
 = \begin{pmatrix}
     1 & -\frac{\sqrt{1-x_c^2}}{x_c} & 0 \\
     0 & 1 & 0 \\
     0 & 0 & 1
    \end{pmatrix}
     \begin{pmatrix}
 U\\V\\W
 \end{pmatrix}, \label{A4}
\end{equation}
where $S$ is a matrix, whose columns are the (generalized) eigenvectors 
corresponding to the eigenvalues of the Jacobian matrix.
From Eq.~(\ref{A4}) it follows that
\begin{equation}
 U=u+\frac{\sqrt{1-x_c^2}}{x_c}v,  \quad
 V=v,  \quad W=w. \label{cv2}
\end{equation}
Note that the phase space on the plane $\{U,V\}$ becomes an ellipse 
lying entirely on the left half-plane $U<0$, for $x_c>0$,
or on the right half-plane $U>0$, for $x_c<0$.
In these new variables the dynamical system becomes
\begin{subequations}
\label{U,V,W-eta-n=1}
\begin{align}
 U_\eta & = \sqrt6\beta x_c U + F_1(U,V,W),
 \label{U-eta-n=1}
\\
 V_\eta & = 3x_c\sqrt{1-x_c^2} \bigg( x_c-\frac{\mu}{\sqrt6} \bigg) W
 +F_2(U,V,W),
\\
 W_\eta & = F_3(U,V,W),
\end{align}
\end{subequations}
where $F_i$ are $\mathcal{O}(U^2,V^2,W^2,UV,UW,VW)$.

Now, the center manifold is given by $W^c=\{(U,V,W)| U=h(V,W), \, h(0,0)=0, \,
\nabla h(0,0)=0 \}$, where $h$, defined on some neighborhood of the critical
point, is a solution of the partial differential equation
\begin{align}
& \frac{\partial h}{\partial V} \left[
 3x_c\sqrt{1-x_c^2} \bigg( x_c-\frac{\mu}{\sqrt6} \bigg) W
 +F_2\Big(h(V,W),V,W\Big) \right] \nonumber
\\
 &\hspace{5mm}+
 \frac{\partial h}{\partial W} F_3\Big(h(V,W),V,W\Big) -
 \sqrt6\beta x_c h(V,W)\nonumber
\\
 &\hspace{5mm} -F_1\Big(h(V,W),V,W\Big)=0.
\end{align}

We search for an order-$m$ solution ($m\geq2$) to this equation of the form
\begin{equation}
 h(V,W)=\sum_{j=2}^m \sum_{i=0}^j a_{i,j-i}V^i W^{j-i},
\end{equation}
where $a_{ij}$ are constants to be determined.
After some algebra, we find that the center manifold is
given by
\begin{align}
 U=h(V,W)
  &=-\frac{1}{2x_c^3}V^2-\frac{\sqrt{1-x_c^2}}{2x_c^5} V^3 \nonumber
\\
  &\hspace{4mm} -\frac{5 - 4 x_c^2}{8 x_c^7} V^4
 +\mathcal{O}(V^5).
 \label{CM}
\end{align}
Note that the coefficients $a_{ij}$ of the terms containing powers of 
$W$ vanish identically.
The flow on the center manifold is determined by the differential equations
\begin{widetext}
\begin{subequations}
\label{flow-V,W}
\begin{align}
V_\eta & = 3x_c\sqrt{1-x_c^2} \bigg( x_c-\frac{\mu}{\sqrt6} \bigg) W
   + \frac{1}{2x_c} \bigg( \sqrt6\mu - 12x_c
   - 2\sqrt6\mu x_c^2 + 18x_c^3  \bigg) V W \nonumber
\\
  & \hspace{4mm} + \frac{\sqrt{1-x_c^2}}{4x_c^3} (\sqrt6\mu 
  + 2\sqrt6\mu x_c^2-36x_c^3) V^2 W 
    -\frac12 x_c\sqrt{1-x_c^2} \bigg( x_c-\frac{\mu}{\sqrt6} \bigg) W^3
    + \frac{1}{4x_c^5} \bigg( \sqrt6\mu - 12 x_c^5\bigg) V^3 W
   \nonumber
\\
  & \hspace{4mm} -\frac{1}{12x_c} \bigg( \sqrt6\mu - 12x_c 
      - 2\sqrt6\mu x_c^2 + 18x_c^3 \bigg) V W^3 
     +\mathcal{O}(V^5,V^4W,V^3W^2,V^2W^3,VW^4,W^5),
 \label{flow-V}
\\
 W_\eta & = \sqrt6 x_c W^3 - \sqrt6 \frac{\sqrt{1-x_c^2}}{x_c} V W^3
 +\mathcal{O}(V^5,V^4W,V^3W^2,V^2W^3,VW^4,W^5).
 \label{flow-W}
\end{align}
\end{subequations}
\end{widetext}

Note that, for $W=0$, $V_\eta$ vanishes; this is expected since the critical
points under consideration belong to a critical line lying on the plane $W=0$
($z=-\pi/2$) and therefore along that line there are no orbits approaching or
moving away from the critical point. Note also that, in the equation for
$V_\eta$, we have to retain terms up to fourth order, since for $x_c=1$ and
$\mu=\sqrt6$ lower-order terms vanish.

Let us now analyze the behavior of the orbits near each point of the critical
line $A(x_c,\sqrt{1-x_c^2},-\pi/2)$, where $x_c\neq0$.
There are three values of $x_c$ that require special consideration.

We start with $x_c=-1$. At lowest order in powers of $V$ and $W$, the center
manifold is given by $U=V^2/2$ and the flow on it is determined by
\begin{subequations}
	\label{xc-1A}
\begin{align}
V_\eta & =3\bigg(1+\frac{\mu}{\sqrt6}\bigg)VW, \\
W_\eta & =-\sqrt6 \, W^3.
\end{align}
\end{subequations}
Taking into account that $\mu>0$ and that, in the neighborhood of the
critical point, $W>0$ and $V>0$, it follows that $V_\eta>0$ and $W_\eta<0$,
implying that the orbits approach the critical point along the
$W$-direction and move away from it along the $V$-direction.
Furthermore, from Eq.~(\ref{U-eta-n=1}), we know that $U_\eta=-\sqrt6\beta U$,
where $U>0$. For $\beta>0$, the orbits approach the critical point along the
$U$-direction, while for $\beta<0$ they move away from it.
Therefore, the critical point $A(-1,0,-\pi/2)$ is a saddle;
the behavior of the orbits near it are depicted in
Figs.~\ref{fig-n1-back}--\ref{fig-n1-front}.

For $x_c=1$, at lowest order in powers of $V$ and $W$, the center manifold is
given by $U=-V^2/2$ and the flow on it is determined by
\begin{subequations}
	\label{xc+1A}
\begin{align}
V_\eta & =3\bigg(1-\frac{\mu}{\sqrt6}\bigg)VW
 -\frac12 \bigg( 1-\frac{\mu}{\sqrt6} \bigg)VW^3 \nonumber
\\
 & \hspace{4mm} -3\bigg( 1-\frac{\mu}{2\sqrt6} \bigg)V^3W,
\\
W_\eta & =\sqrt6 \, W^3,
\end{align}
\end{subequations}
while $U_\eta=\sqrt6\beta U$ gives the flow in the $U$-direction. Taking into
account that, in the neighborhood of the critical point, $W>0$, $V>0$, and
$U<0$, it follows that i) $U_\eta<0$ if $\beta>0$ and $U_\eta>0$ if $\beta<0$
and, consequently, along the $U$-direction, orbits move away or approach the
critical point, respectively; ii) $V_\eta<0$ for $\mu\geq\sqrt6$ and
$V_\eta>0$ for $0<\mu<\sqrt6$ and, consequently, along the $V$-direction,
orbits approach or move away the critical point, respectively; iii) $W_\eta>0$
and, consequently, along the $W$-direction orbits move away from the critical
point. Putting it all together, the critical point $A(1,0,-\pi/2)$ is a
repeller for $\beta>0$ and $0<\mu<\sqrt6$ and a saddle in all other cases
(Figs.~\ref{fig-n1-back}--\ref{fig-n1-front}).

For $x_c=\mu/\sqrt6$ (where $\mu<\sqrt6$), at lowest order in powers of $V$
and $W$, the center manifold is given by $U=-(3\sqrt6/\mu^3)V^2$ and the flow
on it is determined by
\begin{subequations}
	\label{xc-muA}
\begin{align}
V_\eta & =-3\bigg(1-\frac{\mu^2}{6}\bigg) V W,
\\
W_\eta & =\mu W^3,
\end{align}
\end{subequations}
while $U_\eta=\beta\mu U$ gives the flow in the $U$-direction. Taking into
account that, in the neighborhood of the critical point, $W>0$ and $U<0$, it
follows that: i) $U_\eta<0$ if $\beta>0$ and $U_\eta>0$ if $\beta<0$ and,
consequently, along the $U$-direction, orbits move away or approach the
critical point, respectively; ii) $V_\eta<0$ for $V>0$ and $V_\eta>0$ for
$V<0$, implying, consequently, that, along the $V$-direction, orbits approach
the critical point from both sides; iii) $W_\eta>0$ and, consequently, orbits
move away from the critical point along the $W$-direction. Thus, the critical
point $A(\mu/\sqrt6,\sqrt{1-\mu^2/6},-\pi/2)$ is a saddle (see
Figs.~\ref{fig-n1-top-bottom} and \ref{fig-n1-front}).

Let us now consider the case
$x_c \in \left]-1,0\right[\, \cup\, \left]0,\mu/\sqrt6\right[\, \cup\,
\left]\mu/\sqrt6,1\right[$.
Here, according to Eq.~(\ref{flow-V}), the lowest-order term in the expression
for $V_\eta$ is linear and proportional to $W$ ($W>0$). This circumstance
implies a behavior of the orbits crucially different from the behavior
observed for $x_c=\pm 1, \mu/\sqrt6$, where an orbit on the center manifold
with $V(\eta_0)=0$ approaches or moves away from the critical point along the
$W$-direction (vertically on Fig.~\ref{fig-n1-front}) since $V_\eta=0$. 
Therefore, an orbit on the center manifold with $V(\eta_0)=0$
does not approach or move away from the critical point along the
$W$-direction; instead, it drifts along the $V$-direction (see
Fig.~\ref{fig-n1-front}). The direction of this quasi-horizontal drift depends
on the value of $x_c$ and also on the value of $\mu$. Indeed, for
$0<\mu<\sqrt6$, the lowest-order term on the expression for $V_\eta$ is
positive for $x_c\in \left]-1,0\right[ \cup \left]\mu/\sqrt6,+1\right[$
and negative for
$x_c\in \left]0,\mu/\sqrt6\right[$,
implying that, for $x_c\in \left]-1,\mu/\sqrt6\right[$, orbits
are oriented in the direction of growing $x$ and for
$x_c\in \left]\mu/\sqrt6,+1\right[$
orbits are oriented in the opposite direction (see panel (a) of
Fig.~\ref{fig-n1-front}). For $\mu\geq\sqrt6$, the lowest-order term is
positive for $x_c\in \left]-1,0\right[$ and negative for 
$x_c\in \left]0,+1\right[$,
implying that orbits are oriented in the direction of growing $x$
(see panel (b) of Fig.~\ref{fig-n1-front}).
From Eq.~(\ref{flow-W}) we can conclude that for $x_c<0$ the orbits approach
the critical line ($W$ decreases), while for $x_c>0$ they move away
($W$ increases). Near the critical line, this increase or decrease is
very slow (almost unnoticeable in Fig.~\ref{fig-n1-front})
because, at lowest order, $W_\eta\propto W^3$ and $W\ll 1$.

The behavior of the orbits near the critical point 
$A(x_c,\sqrt{1-x_c^2},-\pi/2)$, for $x_c\neq0$ and all possible
values of the parameters $\beta$ and $\mu$ is summarized in
Table.~\ref{Table:criticalpoints}.

\begin{table*}[t]
\begin{tabular}{lllcc}
\hline\hline
& & & $A(x_c,\sqrt{1-x_c^2},-\pi/2)$
    & $B(x_c,\sqrt{1-x_c^2},\pi/2)$ \\
\hline
\multirow{10}{*}{$\beta>0$}
& \multirow{6}{*}{$\quad0<\mu<\sqrt6\quad$} & $x_c=-1$ 
											& A-R-A  & R-R-R  \\
 &  & $x_c\in\left]-1,0\right[$ 						& A-AR-A & R-AR-R \\
 &  & $x_c\in\left]0,\mu/\sqrt6\right[$ 		& R-AR-R & A-AR-A \\
 &  & $x_c=\mu/\sqrt6$ 				& R-A-R  & \textbf{A-A-A}\\
 &  & $x_c\in\left]\mu/\sqrt6,1\right[$		& R-AR-R & A-AR-A \\
 &  & $x_c=1$ 						& R-R-R & A-R-A \\
  \cline{2-5}
& \multirow{4}{*}{$\quad\mu\geq\sqrt6\quad$} & $x_c=-1$ 
								& A-R-A & R-R-R \\ 
 &  & $x_c\in\left]-1,0\right[$			& A-AR-A & R-AR-R \\
 &  & $x_c\in\left]0,1\right[$ 			& R-AR-R & A-AR-A \\
 &  & $x_c=1$					& R-A-R  & \textbf{A-A-A} \\
\hline
\multirow{10}{*}{$\beta<0$}
& \multirow{6}{*}{$\quad0<\mu<\sqrt6\quad$} & $x_c=-1$ 
										& R-R-A  & A-R-R  \\
 &  & $x_c\in\left]-1,0\right[$ 					& R-AR-A & A-AR-R \\
 &  & $x_c\in\left]0,\mu/\sqrt6\right[$ 	& A-AR-R & R-AR-A \\
 &  & $x_c=\mu/\sqrt6$ 					& A-A-R  & R-A-A  \\
 &  & $x_c\in\left]\mu/\sqrt6,1\right[$	& A-AR-R & R-AR-A \\
 &  & $x_c=1$ 							& A-R-R  & R-R-A  \\ \cline{2-5}
& \multirow{4}{*}{$\quad\mu\geq\sqrt6\quad$} & $x_c=-1$ & R-R-A & A-R-R \\ 
 &  & $x_c\in\left]-1,0\right[$			& R-AR-A & A-AR-R\\
 &  & $x_c\in\left]0,1\right[$ 			& A-AR-R & R-AR-A\\
 &  & $x_c=1$					& A-A-R  & R-A-A \\
\hline\hline
\end{tabular}
 \caption{\label{Table:criticalpoints} Behavior of the orbits of the
dynamical system~(\ref{DSn}) near the critical points
$A(x_c,\sqrt{1-x_c^2},-\pi/2)$ and $B(x_c,\sqrt{1-x_c^2},\pi/2)$
for $x_c\neq0$ and all possible values of the parameters $\beta$ and $\mu$.
The letters A, R, and AR in the fourth and fifth columns indicate how the 
orbits behave in different directions. 
The first, second, and third positions of the set of letters
correspond to the $U$, $V$, and $W$ directions, respectively. Furthermore, 
the letter A indicates that the orbit is approaching the critical point, 
the letter R that it moves away, and the letters AR that it approaches on 
one side and moves away on the other.
Thus, the sets A-A-A (highlighted in bold) and R-R-R 
denote attractors and repellers, respectively, 
while the remaining sets denote different types of saddles points.
Note that in this table are not included critical points corresponding to
$x_c=0$, see text in Appendix.}
\end{table*}

Finally, we turn to the case $x_c=0$ and analyze the stability of the
critical point $A(0,1,-\pi/2)$. As mentioned above, in this case all
eigenvalues of the Jacobian matrix $J$ vanish (see Eq.~(\ref{jacA}))
and, consequently, the center manifold theory cannot be applied.
Therefore, we have to use another approach
to analyze the stability of the critical point $A(0,1,-\pi/2)$.

For $\beta<0$, continuity arguments allow us to conclude that,
the critical point $A(0,1,-\pi/2)$ behaves exactly as the 
neighboring critical points;
near it, the orbits are oriented in the direction of growing $x$, 
with $z$ almost constant. From there they turn upward, heading to an 
attracting critical point lying on the plane $z=\pi/2$.

For $\beta>0$, the situation is quite different. Near the surface 
$z=-\pi/2$, orbits quickly spiral to some point $(x,y)=(x_0,y_0)$,
while $z$ remains almost constant; from there, the orbits turn upward,
moving along an approximately straight line away from $A(0,1,-\pi/2)$
(see Fig.~\ref{fig:typical-trajectory-uvw-1}).
In order to see this let us introduce new variables
\begin{equation}
u = x, \qquad v = y - 1, \qquad w = z + \pi/2,  
\end{equation}
bringing the critical point $A(0,1,-\pi/2)$ to the origin,
and define a new parameter $\epsilon = (\sin w)/\beta>0$
(note that $0\leq w\leq\pi$).
For fixed $w$ (positive and small),
we obtain a two-dimensional dynamical system on the 
variables $u$ and $v$, which admits the critical point $(u_0,v_0)$, 
where
\begin{equation}
u_0 = \frac{\sqrt6}{2}\epsilon - \frac{\sqrt6}{2}\mu\epsilon^2 +
\mathcal{O} \left(\epsilon^3\right)
\label{u0-A}
\end{equation}
and
\begin{equation}
v_0 = -\frac{\mu\epsilon}{2} + \frac{3}{8}\left(\mu^2 + 2\right)\epsilon^2
+ \mathcal{O} \left(\epsilon^3\right).
\label{v0-A}
\end{equation}
Note that this critical point is physical, i.e.,
it belongs to the two-dimensional phase space, the upper
half of the unit circle centered at $u=v=0$.
Note also that $(u_0,v_0)\rightarrow(0,0)$ for $\epsilon\rightarrow0$.
Linearizing the two-dimensional dynamical system about the critical point
$(u_0,v_0)$, we obtain
\begin{widetext}
\begin{equation}
 \begin{pmatrix}
	u_\eta \\
	v_\eta
 \end{pmatrix}
=
 \begin{pmatrix}
 -\frac{3}{2}\beta\mu\epsilon^2 + \mathcal{O}\left(\epsilon^3\right) &
 \sqrt6\beta + \frac{\sqrt6}{2}\beta\mu\epsilon
 - \frac{\sqrt6}{8}\beta\left(4\beta^2 + \mu^2 + 6\right)\epsilon^2 +    
 \mathcal{O}\left(\epsilon ^3\right) \\
 -\frac{\sqrt6}{2}\beta\mu\epsilon 
 +\frac{\sqrt6}{4}\beta (\mu^2 +6)\epsilon^2
 +\mathcal{O}\left(\epsilon^3\right) &
 -3\beta\epsilon + 3\beta\mu\epsilon^2 
 + \mathcal{O}\left(\epsilon ^3\right)
\end{pmatrix}
\begin{pmatrix}
u - u_0 \\
v - v_0
\end{pmatrix}.
\label{u,v_eta-A}
\end{equation}
\end{widetext}
The corresponding eigenvalues are
\begin{equation}
 \lambda_{1,2}= -\frac32\beta\epsilon
 \pm i\beta \sqrt{3\mu\epsilon}
 \left(1-\frac{15\epsilon}{8\mu}\right)
 + \mathcal{O}\left(\epsilon ^3\right).
\end{equation}
As $\beta>0$ and $\epsilon>0$, both eigenvalues have negative real part
and a nonvanishing imaginary part (recall that $\mu>0$), implying that the
critical point $(u_0,v_0)$ is an attracting spiral.
The behavior of the variable $w$ is determined by the differential equation
\begin{equation}
 w_\eta=\sqrt6 u \sin^3 w.
\end{equation}
Taking into account that, at linear approximation,
\begin{equation}
 u \approx u_0=\frac{\sqrt6}{2\beta}\sin w
\end{equation}
and $\sin w\approx w$, we conclude that
$w_\eta\approx (3/\beta)w^4$, which, as $\beta>0$, means that the
variable $w$ increases.
Therefore, we conclude that, after rapidly approaching the critical
point $B(0,1,-\pi/2)$,
executing a spiral movement around the point
\begin{equation}
 (x_0,v_0)=\bigg( \frac{\sqrt6}{2}\epsilon,1-\frac{\mu}{2}
 \epsilon \bigg), \qquad
 0<\epsilon\ll 1,
\end{equation}
the orbits are repelled, heading to the attractor lying on the plane
$z=\pi/2$ (see Fig.~\ref{fig:typical-trajectory-uvw-1}).

\begin{figure}[t]
 \includegraphics{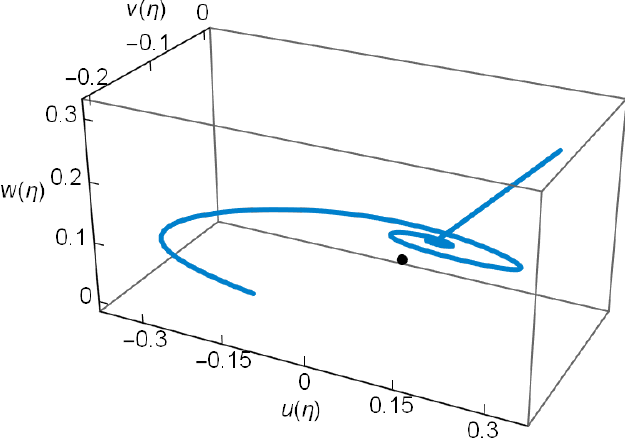}
\caption{\label{fig:typical-trajectory-uvw-1}A typical orbit in the
$\{u,v,w\}$ space near the critical point (0,0,0), for the parameter choices
$\beta=1$ and $\mu=1$. First, the orbit spirals to a point $u=u_0$,
$v=v_0$, with $w$ practically remaining fixed,
and from there it is repelled,
heading to the attractor lying on the plane $w=\pi$.}
\end{figure}

\subsection{Critical line $B(x,\sqrt{1-x^2},\pi/2)$ \label{Sec-B}}

The analysis of the critical line $B(x,\sqrt{1-x^2},\pi/2)$ is analogous
to the one in the previous subsection for the critical line 
$A(x,\sqrt{1-x^2},-\pi/2)$,
but with some crucial differences due to changes of sign.

Let us consider a specific critical point $B(x_c,\sqrt{1-x_c^2},\pi/2)$,
where $x_c\in \left[-1,1\right]$.
The Jacobian matrix of the dynamical system~(\ref{DSn}) is given by
Eq.~(\ref{jacA}) with switched sign on the right-hand side
and the corresponding eigenvalues are $\lambda_1=-\sqrt6\beta x_c$,
$\lambda_2=0$, and $\lambda_3=0$.
Expanding $\sin z$ and $\cos z$ in the dynamical system (\ref{DSn})
in a Taylor series about $z=\pi/2$ and introducing new variables
\begin{equation}
 u=x-x_c, \quad v=y-\sqrt{1-x_c^2}, \quad w=z-\pi/2,
\end{equation}
which shift the critical point $B(x_c,\sqrt{1-x_c^2},\pi/2)$ to the origin,
the system takes the form given by Eqs.~(\ref{u,v,w-eta-n=1})
with switched signs on the right-hand sides.

A second change of variables (for $x_c\neq0$),
given by Eq.~(\ref{cv2}), brings the dynamical system to the form given by 
Eqs.~(\ref{U,V,W-eta-n=1}) with switched signs on the right-hand sides,
which permits direct application of the center manifold theorem.
The center manifold is given by Eq.~(\ref{CM})
and the flow on it is determined by the differential equations 
(\ref{flow-V,W}) with switched signs for all terms on the right-hand sides.

For $x_c=-1$, at lowest order in powers of $V$ and $W$, the center manifold
is given by $U=V^2/2$ and the flow on it is determined by 
Eqs.~(\ref{xc-1A}) with switched signs on the right-hand sides,
while $U_\eta=\sqrt6\beta U$ gives the flow in the $U$-direction. Taking into
account that, in the neighborhood of the critical point, $U>0$, $V>0$, and
$W<0$, it follows that $V_\eta>0$ and $W_\eta<0$, implying that the orbits
move away from the critical point along the $W$ and $V$ directions.
Furthermore, for $\beta>0$, the orbits move away the critical point along the
$U$-direction, while for $\beta<0$ they approach it. Therefore, for $\beta>0$
the critical point $B(-1,0,\pi/2)$ is a repeller and for $\beta<0$ is a saddle
(see Figs.~\ref{fig-n1-back}--\ref{fig-n1-front}).

For $x_c=1$, at lowest order in powers of $V$ and $W$, the center manifold
is given by $U=-V^2/2$ and the flow on it is determined by
Eqs.~(\ref{xc+1A}) with switched signs on the right-hand sides,
while $U_\eta=-\sqrt6\beta U$ gives the flow in the $U$-direction. Taking into
account that, in the neighborhood of the critical point, $U<0$, $V>0$, and
$W<0$, it follows that i) $U_\eta>0$ if $\beta>0$ and $U_\eta<0$ if $\beta<0$
and, consequently, along the $U$-direction, orbits approach or move away from
the critical point, respectively; ii) $V_\eta<0$ for $\mu\geq\sqrt6$ and
$V_\eta>0$ for $0<\mu<\sqrt6$ and, consequently, along the $V$-direction,
orbits approach or move away the critical point, respectively; iii) $W_\eta>0$
and, consequently, along the $W$-direction orbits approach the critical point.
Putting it all together, the critical point $B(1,0,\pi/2)$ is an attractor for
$\beta>0$ and $\mu\geq\sqrt6$ and a saddle in all other cases (see
Figs.~\ref{fig-n1-back}--\ref{fig-n1-front}).

For $x_c=\mu/\sqrt6$, with $\mu<\sqrt6$, at lowest order in powers of $V$
and $W$, the center manifold is given by $U=-(3\sqrt6/\mu^3)V^2$ and the
flow on it is determined by
Eqs.~(\ref{xc-muA}) with switched signs on the right-hand sides,
while $U_\eta=-\beta\mu U$ gives the flow in the $U$-direction. Taking into
account that, in the neighborhood of the critical point, $U<0$ and $W<0$, it
follows that: i) $U_\eta>0$ if $\beta>0$ and $U_\eta<0$ if $\beta<0$ and,
consequently, along the $U$-direction, orbits approach or move away from the
critical point, respectively; ii) $V_\eta<0$ for $V>0$ and $V_\eta>0$ for
$V<0$, implying, consequently, that, along the $V$-direction, orbits approach
the critical point from both sides; iii) $W_\eta>0$ and, consequently, orbits
approach the critical point along the $W$-direction. Thus, the critical point
$B(\mu/\sqrt6,\sqrt{1-\mu^2/6},\pi/2)$ is an attractor for $\beta>0$ and
a saddle for $\beta<0$ (see Figs.~\ref{fig-n1-top-bottom} and
\ref{fig-n1-front}).

For $x_c\neq0,\pm1,\mu/\sqrt6$, the lowest-order term on the expression for
$V_\eta$ is linear and proportional to $W$ ($W>0$), implying that orbits on
the center manifold with $V(\eta_0)=0$ do not approach or move away from the
critical point along the $W$-direction, but, instead, they drift along the
$V$-direction (see Fig.~\ref{fig-n1-front}). For $0<\mu<\sqrt6$, orbits are
oriented in the direction of growing $x$ for
$x_c\in \left]-1,\mu/\sqrt6\right[$ and in
the opposite direction for $x_c\in \left]\mu/\sqrt6,+1\right[$
(see panel (a) of Fig.~\ref{fig-n1-front});
for $\mu\geq\sqrt6$, orbits are oriented in the direction of growing $x$
(see panel (b) of Fig.~\ref{fig-n1-front}).
Furthermore, for $x_c<0$, orbits approach the critical line ($W$ decreases),
while for $x_c>0$ they move away ($W$ increases); however, near the critical
line, this increase or decrease is almost unnoticeable, since, at lowest
order, $W_\eta\propto W^3$ and $W\ll 1$.

Finally, let us analyze the case $x_c=0$. Since all eigenvalues of the
Jacobian matrix vanishes, we can not use the center manifold
theorem. Therefore, to analyze the stability of the critical point 
$B(0,1,\pi/2)$ we use another approach.

For $\beta>0$, continuity arguments allow us to conclude that,
similarly to the neighboring critical points,
the critical point $B(0,1,\pi/2)$ is unstable; 
near it, the orbits are oriented in the direction of growing $x$,
heading to the attractor $B(\mu/\sqrt6,\sqrt{1-\mu^2/6},\pi/2)$.

For $\beta<0$, the situation is quite different; the critical
point $B(0,1,\pi/2)$ is an attractor. 
In Fig.~\ref{fig:typical-trajectory-uvw-2} a typical trajectory
is plotted in the shifted variables
\begin{equation}
u = x\>, \qquad v = y - 1\>, \qquad w = z - \pi/2\>. 
  \label{uvw}
\end{equation}
What is clearly visible is the fact that the $u$ and $v$ variables
quickly spiral to a point $u=u_0$, $v=v_0$, with $w$ practically
remaining fixed. 
From there on, the orbit continues in a practically straight line toward
the critical point $u=v=w=0$.
The reason this happens is that the approach of the $u$, $v$ variables to
the point $(u_0,v_0)$ is exponentially fast, while the approach of the $w$
variable toward $0$ goes like $w(\eta) \propto -\eta^{-1/3}$ as
$\eta \to \infty$.

In order to see this, it is useful to linearize Eqs.~(\ref{DSn-x}) 
and (\ref{DSn-y}), for fixed $w$.
Let us define  the parameter $\epsilon = -(\sin w)/\beta$,
which has the same sign as $\beta$ (note that $w$ is taken to have small
negative values).
It follows that the relevant linearization yields
\begin{widetext}
\begin{equation}
 \begin{pmatrix}
	u_\eta \\
	v_\eta
 \end{pmatrix}
=
 \begin{pmatrix}
 \frac{3}{2}\beta\mu\epsilon^2 + \mathcal{O}\left(\epsilon^3\right) &
 -\sqrt6\beta + \frac{\sqrt6}{2}\beta\mu\epsilon
 + \frac{\sqrt6}{8}\beta\left(4\beta^2 + \mu^2 + 6\right)\epsilon^2 +    
 \mathcal{O}\left(\epsilon ^3\right) \\
 -\frac{\sqrt6}{2}\beta\mu\epsilon 
 - \frac{\sqrt6}{4}\beta (\mu^2 +6)\epsilon^2
 + \mathcal{O}\left(\epsilon^3\right) &
 -3\beta\epsilon - 3\beta\mu\epsilon^2 
 + \mathcal{O}\left(\epsilon ^3\right)
\end{pmatrix}
\begin{pmatrix}
u - u_0 \\
v - v_0
\end{pmatrix}
\label{u,v_eta}
\end{equation}
\end{widetext}
with
\begin{equation}
u_0 = -\frac{\sqrt6}{2}\epsilon - \frac{\sqrt6}{2}\mu\epsilon^2
+ \mathcal{O} \left( \epsilon^3 \right) \label{u0}
\end{equation}
and
\begin{equation}
v_0 = \frac{\mu\epsilon}{2} + \frac38 \left( \mu^2 + 2 \right) \epsilon^2
+ \mathcal{O} \left( \epsilon^3 \right). \label{v0}
\end{equation}

\begin{figure}[t]
 \includegraphics{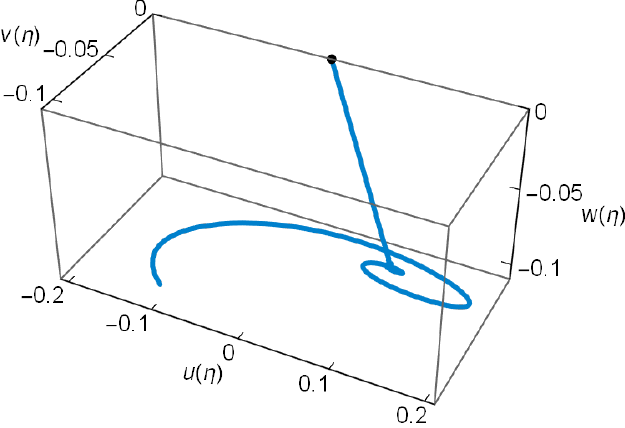}
\caption{\label{fig:typical-trajectory-uvw-2}A typical trajectory in the
$\{u,v,w\}$ space toward the critical point $(0,0,0)$, for the parameter 
choices $\beta = -1$, $\mu = 1$.
Clearly visible is the exponentially fast convergence of the variables
$u$ and  $v$ toward the values $u_0$ and $v_0$ at almost fixed value of $w$,
and the continuation, from there on, in a practically straight line
toward the critical point.}
\end{figure}

Note that for $\beta>0$ (and thus $\epsilon>0$) the point $(u_0,v_0)$ does 
not belong to the phase space and, therefore, the behavior depicted in 
Fig.~\ref{fig:typical-trajectory-uvw-2} does not take place; as mentioned
above, in this case, the orbits pass near the critical point
$B(0,1,\pi/2)$, heading to the attractor
$B(\mu/\sqrt6,\sqrt{1-\mu^2/6},\pi/2)$.

The eigenvalues of the matrix on the right-hand side of Eq.~(\ref{u,v_eta})
are
\begin{equation}
 \lambda_{1,2}= -\frac32\beta\epsilon
 \pm i\beta \sqrt{-3\mu\epsilon}
 \left(1+\frac{15\epsilon}{8\mu}\right)
 + \mathcal{O}\left(\epsilon^3\right).
\end{equation}
As $\beta<0$ and $\epsilon<0$, both eigenvalues have negative real part
and a nonvanishing imaginary part (recall that $\mu>0$), implying that the
critical point $(u_0,v_0)$ is an attracting spiral.

It is now easy to determine the time development of the third variable $w$.
After the initial phase in which the trajectory converges for approximately
fixed value of $w = -\arcsin(\epsilon\beta)$ to the values 
$u = u_0$ and $v = v_0$ given by Eqs.~(\ref{u0})
and (\ref{v0}), $w$ will increase to the asymptotic value 0.
Meanwhile, the values of $u$ and $v$ will shift slowly,
remaining in very good approximation equal to the slowly shifting values
$u_0\bigl(\epsilon \to -(\sin w)/\beta\bigr)$
and $v_0\bigl(\epsilon \to -(\sin w)/\beta\bigr)$, respectively.
From Eq.~(\ref{DSn-z}) it follows that
\begin{align}
w_\eta & \approx \sqrt{6}u_0\bigl[\epsilon = -(\sin w)/\beta\bigr]
\bigl(\cos(\pi/2 + w)\bigr)^3 \nonumber
\\
& = -\frac{3}{\beta} w^4 + \mathcal{O}(w^5).
\end{align}
The asymptotic solution for $\eta \to\infty$ is
\begin{equation}
w(\eta) \sim -\sqrt[3]{\frac{-\beta}{\eta}}.
\label{w-eta}
\end{equation}
Indeed $w$ (which is negative) approaches zero very slowly
(note that $\beta$ is assumed to be negative).

The evolution (\ref{w-eta}) contrasts sharply with the
exponential approach of $(u,v)$ toward the point $(u_0,v_0)$
at constant $w$
determined by Eq.~(\ref{u,v_eta}).
The latter amounts to an elliptic spiral
of which the major and minor axis both are proportional to
$ e^{-\lambda \eta} $, where
\begin{equation}
\lambda = \frac{3}{2}\beta\epsilon + \mathcal{O}(\epsilon^2)
= -\frac{3}{2}\sin w + \mathcal{O}\bigl((\sin w)^2\bigr) > 0
\end{equation}
equals minus the real part of the (complex) eigenvalues of the matrix at the
right-hand side of Eq.~(\ref{u,v_eta}) (which are equal).

\subsection{Critical point $C(0,0,0$)\label{Sec-C}}

Let us now consider the critical point $C(0,0,0)$. The Jacobian matrix of the
dynamical system~(\ref{DSn}),
\begin{equation}
 J=\begin{pmatrix}
     -\frac32 & 0 & \frac{\sqrt6 }{2}\beta\\
     0 & \frac32 &0 \\
     \sqrt6 &0& 0
    \end{pmatrix},
\end{equation}
yields the eigenvalues
\begin{equation}
 \lambda_1=\frac32, \quad \lambda_{2,3}=
 -\frac34 \left( 1\pm\sqrt{1+\frac{16}{3}\beta} \right)
\end{equation}
and the corresponding eigenvectors
\begin{equation}
 \vec{v}_1=\left( \begin{matrix}
 0 \\ 1 \\ 0
 \end{matrix} \right),
 \quad 
 \vec{v}_{2,3}=\left( \begin{matrix}
 -\frac{\sqrt6}{8} \left( 1\pm \sqrt{1+\frac{16}{3}\beta}
   \right) \\ 0 \\ 1
 \end{matrix} \right).
\end{equation}

Because all eigenvalues are different from zero (recall that $\beta\neq0)$,
the linear theory suffices to
analyze the stability of this critical point.

Here, we have to consider three cases.

First, the case $\beta>0$. Since $\lambda_1>0$, $\lambda_2<0$, and
$\lambda_3>0$, the critical point $C$ is a saddle; the orbits move away from
it along the $y$-direction and the direction defined by the third eigenvector,
and approach it along the direction defined by the second eigenvector (see
panel (a) of Fig.~\ref{fig-n1-back}).

Second, the case $-3/16\leq\beta<0$. Since $\lambda_1>0$, $\lambda_2<0$, and
$\lambda_3<0$, the critical point $C$ is again a saddle (an attracting node in
the plane $y=0$); the orbits move away from it along the $y$-direction and
approach it along the directions defined by $\vec{v}_2$ and $\vec{v}_3$
(see panel (b) of Fig.~\ref{fig-n1-back}).

Third, the case $\beta<-3/16$, for which $\lambda_2$ and $\lambda_3$ become
complex numbers with negative real part. The critical point is again a saddle
(an attracting spiral in the $y=0$ plane); the orbits move away from it along
the $y$-direction and spiral to it in the $y=0$ plane (see panel (c) of
Fig.~\ref{fig-n1-back}).

\end{document}